\documentclass[3p,preprint,hyperref]{elsarticle}
\input{prolegomenaPC}
\pdfoutput=1

\begin{document}

\begin{frontmatter}
\title{Peaks and cusps: anomalous thresholds and LHC physics}

\author[torino]{Giampiero Passarino}
\ead{giampiero@to.infn.it}

\address[torino]{\csumb}


\begin{abstract}
The behavior of scattering amplitudes in the vicinity of a physical-region Landau singularity is considered.
The impact on LHC processes is discussed.
\end{abstract}

\end{frontmatter}


\section{Introduction \label{intro}}
Resonances are normally observed as peaks in certain invariant mass distributions; however, a question 
arises: is a peak necessarily due to the presence of a resonance? Are there peaks produced by kinematic 
singularities? 

Landau equations for a given Feynman integral are a set of kinematic constraints that are necessary for the appearance 
of a pole or branch point in the integrated function (as a function of external kinematics and masses).
Landau equations admit many families of solutions which are naturally classified as leading Landau singularities
(LLS), sub-leading Landau singularities (SLLS), sub-sub-leading (S${}^2$LLS) \etc  

Leading Landau singularities have been studied mostly in the context of hadron spectroscopy~\cite{Guo:2017wzr} where,
in order to establish an unambiguous strategy it is important to distinguish kinematic singularities from genuine 
resonances, \ie poles of the $\mrS\,$-matrix. An example is $\mrB^+ \to \mrB^0_{\mrs} \pi^+ \pi^0$ via 
$\mrB {\overline{\mrK}}^*$ rescattering~\cite{Liu:2017vsf,Liu:2015taa}.

As an example, a triangle singularity is a logarithmic branch point, which would produce an infinite reaction rate if 
it appears in the physical region. This never happens because at least one of the three particles must be unstable.  
The finite width, introduced by the complex-mass scheme~\cite{Denner:2006ic,Actis:2006rb,Denner:2014zga}, moves the 
singularity into the complex plane, and the 
differential reaction rate can have a finite peak due to the proximity of the singularity. Here, by complex plane we mean
complex plane for Mandelstam invariants or complex hypersurface in the parametric representation of the
corresponding diagram, \ie no pinch singularity in Feynman parameter space.

The subject of Feynman amplitudes with variable momenta and non-zero masses has been studied by physicists since 
the 1950’s, \eg the inelastic scattering $\PKp + \Pp \to \PGpp + \PGpz + \Pp$ (see Figs.[1,2] of
\Bref{RevModPhys.33.448}).
Meanwhile new mathematical methods involving Hodge structures and variations of Hodge structures have been
developed. The use of these techniques to the study of amplitudes and Landau singularities in momentum space has
been described in \Brefs{Bloch:2010gk,Kreimer:2016tqq}.
A determination of the complete set of branch points of amplitudes in planar $N=4$ super-Yang-Mills theory directly 
from the amplituhedron, without resorting to any particular representation in terms of local Feynman integrals has
been presented in \Bref{Dennen:2016mdk}.

However, in recent years, not much attention has been paid to the problem in the context of high energy physics, with the 
noticeable exception of the work by Boudjema and collaborators~\cite{Boudjema:2008zn,LeDuc:2008zz,Do:2010zz}
(see also section~4.4 of \Bref{Denner:1996ug}, \Bref{Binoth:2002xh} and developments in the Golem95 project, \eg 
\Bref{Rodgers:2012maa}). 

In this paper we analyze typical LHC processes looking for possible effects due to the presence of a leading Landau
singularity, or anomalous threshold (hereafter AT).
It should be stressed that by ``cusp'' we mean a cusp in some differential distribution (\eg invariant mass or 
$p_{\mrT}$) and not cusp of a Landau variety~\footnote{A Landau variety $\mrL$ is a reducible algebraic variety; the 
singular points of a generic plane section of $\mrL$ are expected to be transverse intersections, tacnodes or 
cusps~\cite{Ponzano:1970ch}.}.

In the rest of this paper we discuss the complications that arise in dealing with the singular part of a scattering 
amplitudes and the impact of anomalous thresholds on LHC physics.
We begin in \autoref{LLS} by reviewing briefly the definition of leading Landau singularities, illustrating
the introduction of complex masses.

In \autoref{bubbles} -- \autoref{hexa} we describe LLS for two-, three-, $\dots$ six-point 
functions using the complex mass scheme discussed in \autoref{Cmass}.    
In \autoref{tloops} we include two-loop diagrams into our analysis.
In \autoref{Scas} we discuss special configurations leading to non-integrable LLS, even within the regularization 
introduced by complex masses.
QED/QCD induced LLS, not to be mistaken for the infrared ones, are presented in \autoref{nsing}.
In \autoref{BSM} we discuss relatively simple examples of beyond-standard-model (BSM) LLS.
In \ref{TD} -- \ref{SB} technical details are presented.
\section{Leading Landau singularities \label{LLS}}
Learning from the study of singularities of scattering amplitudes:

\bei

\item[\dnuma] For a given Feynman diagram there exists a discriminant 
\bq
\mrD\lpar \spro{p_i}{p_j},m^2_i,\alpha_i \rpar,
\eq
which is an homogeneous polynomial in the $\alpha_i$ and whose coefficients 
are linear in the $\spro{p_i}{p_j}$~\footnote{In our metric, space-like $p$ corresponds to positive $p^2$. Further
$p_4 = i\,p_0$ with $p_0$ real for a physical four-momentum.} and $m^2_i$, such that the equations
$\partial \mrD/\partial \alpha_i = 0$ are equivalent to the usual Landau conditions
for the existence of the singularity, as described in \Bref{Landau:1959fi}. As it is well known, given any $m$ 
homogeneous polynomials in $m$ unknowns there exists a unique minimal homogeneous polynomial in the coefficients
($\mrR$ the resultant) such that $\mrR = 0$ is a necessary and sufficient condition for the existence of a
solution to the system of equations (Landau-Nakanishi equations), distinct from the trivial solution
$\alpha_1 = \,\dots\,\alpha_m = 0$~\footnote{Usually Landau-Nakanishi~\cite{Nakanishi:1968hu,doi:10.1143/PTP.22.128} 
equations are called Landau equations for short, see also Bjorken (thesis, Stanford Univ. $1959$) and 
\Brefs{Taylor:1960zz,Polkinghorne1960,Bloxham:1969cm,Bloxham:1969cp,Bloxham:1969cq,mathematical1993encyclopedic}}. 
Note that $\alpha_i \ge 0$ is required for the scattering to be physical (the so-called $+\,\alpha$ Landau surfaces,
as opposite to ``mixed-$\,\alpha$'' solutions).

\item[\dnumb] The leading Landau singularity requires all of the$\alpha_i$'s to be non zero; the case where some of the 
parameters vanish can be interpreted as the leading singularity of a diagram obtained from the original one
contracting the lines associated to the vanishing $\alpha$'s. Note that the definition used in \Bref{Prlina:2018ukf}
differs slightly from our conventional usage.
Furthermore, for a given set of values $( p_1,\,\dots\,,p_{\ssN} )$ which lie on the given physical-region Landau 
singularity there exists only one unique set of values for the internal momenta which satisfy the Landau
equations.

\eei

For the discussion of LL singularities there are three important theorems.

\begin{theorem}[Coleman and Norton~\cite{Coleman:1965xm}] \label{CNT}
A Feynman amplitude has singularities on the physical boundary if and only if the relevant Feynman diagram can 
be interpreted as a picture of an energy- and momentum-conserving process occurring in space-time, with all internal 
particles real, on the mass shell, and moving forward in time. As a by-product, the Feynman parameter associated 
with an internal line is identified (within a proportionality factor) with the time the particle exists between 
collisions, divided by its mass. 
\end{theorem}

Additional results can be found in 
\Brefs{PhysRev.119.1763,Aitchison:1964zz,Norton:1964zz,1964PhRv..134..687B,MR0129319,PhysRev.132.2703}.
Other results can be found in \Bref{PhysRev.154.1363}, where it is stressed that the meaning of 
``physical boundary'' in the Coleman-Norton theorem is as follows: consider the triangle of 
\refig{ATfig1}, 
the unitary cut starts at $(m_1 + m_3)^2$ (right-hand branch cut); the physical boundary is just above the cut.
To give an example, the original Peierls mechanism~\cite{Peierls:1961zz} gives singularities on the wrong sheet;
the modified Peierls mechanism~\cite{Goebel:1964zz} gives the (triangular) singularity on the correct sheet but it 
has been shown that does not produce peaks in invariant-mass plots, the so-called Schmid theorem~\cite{PhysRev.154.1363}. 
Nevertheless, in \Bref{Goebel:1982yb} it was argued that terms involving the singularity of a triangle diagram can in
principle at least lead to observable effects in the differential cross section.
Other counterexamples can be found in \Bref{Anisovich:1995ab} and in \Brefs{PhysRev.160.1346,Valuev:1977ej}. 
However, one should be aware that most of these papers limit their analysis to triangular singularities.

\begin{theorem}[Kershaw~\cite{Kershaw:1971rc}] \label{KT}
The singular part of a scattering amplitude around its leading Landau 
singularity may be written as an algebraic product of the scattering
amplitudes for each vertex of the corresponding Landau graph times a 
certain explicitly determined singularity factor which depends only on the
type of singularity (triangle graph, box graph, etc.) and on the masses and
spins of the internal particles.
\end{theorem}

It is worth noting that the consequences of the theorem have been reinterpreted by various authors in terms of multiple 
cuts on Feynman 
diagrams~\cite{Ossola:2006us,Bern:1994zx,Berger:2006ci,Bern:2005hh,Britto:2006sj,Ellis:2006ss,Anastasiou:2006jv,Cachazo:2008vp}.

Kershaw theorem is based on the fact that there always exists a finite polynomial
in the scalar product of the external momenta such that $\mrP(\spro{p_i}{p_j}) =
0$ gives the location of the leading Landau singularity. The proof of this
property is particularly simple for one-loop diagrams. 
Consider a scalar, one-loop, $\mrN$-point functions in $\mrd$ dimensions
($\mrd= 4-\ep$): external momenta will be labelled as $p_1,\,\dots\,p_{\ssN}$ and
let us consider ${\cal P}_{\ssN}$ the set of the non-cyclic permutation of 
$\lpar 1,\,\dots\,\mrN \rpar$ with the first entry fixed. Vectors $k_i$ are 
introduced according to the following convention:
\bei
\item[$\mrN=3$] there are two elements, \ie $\lpar 1,2,3 \rpar$ and 
$\lpar 1,3,2 \rpar$. We define
\bqa
\lpar 1,2,3\rpar &\to& \quad k_1 = p_1, \; k_2 = p_2,
\nl
\lpar 1,3,2\rpar &\to& \quad k_1 = p_1, \; k_2 = p_3.
\eqa
\item[$\mrN=4$] There are three elements and we define
\bqa
\lpar 1,2,3,4 \rpar &\to& \quad k_1= p_1, \; k_2= p_2, \; k_3= p_3,
\nl
\lpar 1,2,4,3 \rpar &\to& \quad k_1= p_1, \; k_2= p_2, \; k_3= p_4,
\nl
\lpar 1,3,2,4 \rpar &\to& \quad k_1= p_1, \; k_2= p_3, \;k_3= p_2.,
\eqa
\item[$\mrN=5$] There are twelve elements, \etc
\eei
With the notations of \Bref{Ferroglia:2002mz} we define a scalar integral as
\bq
\mrS_{\mrd\,;\,\ssN}\lpar \mrw \rpar = 
\frac{\mu^{\ep}}{i\,\pi^2}\,\int\,d^{\mrd}q\,
\frac{1}{\prod_{i=0,\ssN-1}\,[i]},
\qquad
[i] = \lpar q + k_0 + \,\cdots\,+ k_i\rpar^2 + m^2_i,
\eq
where $k_0 = 0$ and where $\mrw$ is an element of ${\cal P}_{\ssN}$. Furthermore $\mu$ is the 't Hooft scale and
$\ep= \mrd - 4$. 
In this way we can characterize the whole set of $\mrN\,$-point functions
contributing to a given amplitude and not just one specific diagram.

In parametric space we obtain 
\bq
\mrS_{\mrd\,\ssN}\lpar \mrw \rpar  = \lpar \frac{\mu^2}{\pi}\rpar^{2-\mrd/2}\,
\egam{\mrN-\frac{\mrd}{2}}\,[\mrN]_{\mrd}(\mrw),
\eq
where from the triangle to the hexagon we will use the following notations: $[3] \equiv \mrC,\,\dots\,[7] \equiv \mrG$. 
We have
\bq
[\mrN]_{\mrd}(w) = \dsimp{\mrN-1}\,\mrV^{\mrd/2-\mrN}_{\ssN}(\mrw),
\eq
\bq
\mrV_{\ssN}(w) = x^t\,\mrH_{\ssN}(\mrw)\,x + 2\,\mrK^t_{\ssN}(\mrw)\,x + \mrL_{\ssN}(w),
\quad
X_{\ssN}(w) = - \mrK^t_{\ssN}(\mrw)\,\mrH^{-1}_{\ssN}(\mrw), 
\label{Xdef}
\eq
\bq
\Delta_{\ssN}(w) = \mrL_{\ssN}(\mrw) - \mrK^t_{\ssN}(\mrw)\,\mrH^{-1}_{\ssN}(\mrw)\,\mrK_{\ssN}(\mrw),
\label{Gdef}
\eq
where $\mrH_{ij}= -\,\spro{k_i}{k_j}$; $\mrG_{\ssN} = \mathrm{det}\,\mrH_{\ssN}$ is the Gram determinant associated 
with the $\mrN\,$-point function of argument $\mrw \in {\cal P}_{\ssN}$. Furthermore,
$\mrK_i = - 1/2\,(m^2_i - m^2_{i+1} - k^2_i - 2\,\sum_{j=1,i-1}\,\spro{k_j}{k_i})$, $\mrL= m^2_1$ and
\bq
\dsimp{\mrN-1} = \int_0^1 dx_1\,\int_0^{x_1} dx_2\,\dots\,\int_0^{x_{\ssN-2}} dx_{\ssN - 1} \spp 
\eq
Let $\mrM_{\ssN}$ be the $\mrN\,\times\,\mrN$ matrix
\[
\mrM_{\ssN} = \left(
\begin{array}{cc}
\mrH_{\ssN}   & \mrK_{\ssN} \\
\mrK^t_{\ssN} & \mrL_{\ssN} \\
\end{array}
\right)
\]
Then one can easily prove that
\bq
\Delta_{\ssN}(\mrw) = \frac{\mrC_{\ssN}(\mrw)}{\mrG_{\ssN}(\mrw)} \spc
\qquad 
X^i_{\ssN} = \frac{\mathrm{det}\,\mrM_{(i\,,\,\ssN)}}{\mrG_{\ssN}} \spc
\label{BSTf}
\eq
where $\mrC_{\ssN} = \mathrm{det}\, \mrM_{\ssN}$ is the so-called modified Cayley 
determinant~\cite{Melrose:1965kb,Regge1964} of the diagram~\footnote{The more familiar definition is as follows: 
define propagators $[i] = (q + q_i)^2 + m^2_i$, with $q_0= 0$; introduce the 
matrix $Y_{ij} = 1/2((q_i - q_j)^2 + m^2_i + m^2_j)$ and define the modified Cayley determinant as $\mrC = 
\mathrm{det}\, Y$. To be more precise $\mrC$ is proportional to a signed minor~\cite{Fleischer:2011bi} of the 
modified Cayley determinant defined in \Bref{Regge1964,Melrose:1965kb}.} and we can write
\bq
\mrV_{\ssN} = \lpar x - X_{\ssN} \rpar^t\,\mrH_{\ssN}\,\lpar x - X_{\ssN} \rpar + \Delta_{\ssN} =
y^t\,\mrM_{\ssN}\,y 
\spc
\label{xpar}
\eq
where $y^t= (x^t\,,\,1)$.

\begin{theorem}[Ferroglia et al. ~\cite{Ferroglia:2002mz}; Goria and Passarino~\cite{Goria:2008ny}]
It is easily seen~\footnote{In general for a hypersurface $\mrV(x_1\,\,\dots\,,\,x_n) = 0$
the singular points are those at which all the partial derivatives simultaneously vanish. The notion of singular 
points is a purely local property.
The determination of the multiplicity of a singular point, is based on ascertaining which of the higher-order 
derivatives vanish at that point~\cite{SAKKALIS1990405}.} that $\Delta_{\ssN} = 0$ induces a 
pinch~\cite{Eden:1966dnq,hwa1966homology,Speer:1970ss} on the integration contour at the point $x= X_{\ssN}$; therefore, if
\bq
\Delta_{\ssN} = 0, \qquad 0 < X_{\ssN,\ssN-1} < \,\dots\,< X_{\ssN,1} < 1,
\eq
we have the leading singularity (from \eqn{xpar} we derive that it represents a singular point of multiplicity two). 
Leading singularities of diagrams obtained by shrinking one (or more) line of the original diagram to a point give 
the sub-leading singularities. 

\end{theorem}

In the Cayley language the Landau equations for a general case can be written as follows. Consider the integral
\bq
\mrI = \int_{\mrD}\,\prod_{i=1}^n\,dx_i\,
\mrV^{-\mu}_{\ssN,n}\lpar x_1\,\dots\,x_n\,;\,w_1\,\dots\,w_k\rpar \spc
\label{LS}
\eq
where $\mrV$ is a multivariate polynomial of degree $\ssN$ and an algebraic function of 
$k\,$-parameters $w_1\,\dots\,w_k$. Therefore $\mrV \in \Cf\lrbr x\,,\,w\rrbr$.
$\mrD$ is the domain of integration.
$\mrV_{\ssN,n} = 0$ is the locus of the singularities of the integrand; let 
$\mcB_j\lpar x\,,\,w\rpar$, $j = 1,\,\dots\,,d$ be the set representing the boundary of
$\mrD$.
\begin{proposition}
The necessary conditions for the leading singularities to occur when the hypercontour is pinched 
between the surfaces of singularity or meets a boundary variety are 
\bei
\item[\snitem] $\exists\;\; \alpha_i\,,\,\beta$, not all equal to zero and such that at the point
$w_k = w^0_k$ and $x_k = x^0_k$ we have $\beta\,\mrV_{\ssN,n} = 0$ and
\eei
\bqa
{}&{}& \alpha_i\,\mcB_i\lpar x_i\,,\,w_i\rpar = 0, \qquad i = 1,\,\dots\,,d \spc
\nl
{}&{}& \frac{\partial}{\partial x_i}\,\lrbr \sum_i\,\alpha_i\,\mcB_i\lpar x\,,\,w\rpar + 
\beta\,\mrV_{\ssN,n}\lpar x\,,\,w\rpar \rrbr = 0, \qquad i = 1,\,\dots\,,n \spp
\label{mpinch}
\eqa
\end{proposition}
If $\mrD$ is a $n\,$-dimensional hypercube and 
\bq
\mrV_{\ssN\,,\,n} = \sum_{i=0}^{\ssN}\,\mrV^i_n \spc 
\qquad
\mrV^i_n = \sum_{0 \le i_1 = \,\cdots\,i_n \le i}\,a^i_{i_1\,\dots\,i_n}\,
x^{i_1}_1\,\cdots\,x^{i_n}_n \spc
\eq
where the $\mrV^i_n$ are homogeneous polynomials and $\mrV_{\ssN\,,\,n}$ is a generic polynomial in the
ring of polynomials of degree $\mrN$, it is convenient to determine the $(\ssN - 1)^n$ n-tuples 
$X^i_1\,\dots\,X^i_n$ such that
\bq
\mrV_{\ssN\,,\,n}\lpar x_1-X^i_1\,\dots\,x_n-X^i_n\rpar = 
\Delta + \sum_{i=2}^{\ssN}\,\mrV^i_n\lpar x_1-X^i_1\,\dots\,x_n-X^i_n\rpar,
\quad
i = 1\,\dots\,(\ssN - 1)^n \spc
\label{inv}
\eq
so that the solutions of $\Delta\lpar w_1\,\dots\,w_k\rpar = 0$, are the potential 
(leading) pinch singularities if $X^i_j \in \Rf\,,\;\; 0 < X^i_j < 1 \;\;\forall j$. 
For $\mrV^2_n =\,\dots\,= \mrV^k_n = 0$ the singular point will have multiplicity $k + 1$.

To summarize: a general understanding of the behavior of Feynman amplitudes may be obtained by analogy with the 
behavior of a function $f$ of a single complex variable $\mrz$, defined as a contour integral with
respect to a second variable $s$ of a function $g$ analytic in $\mrz$ and $s$. The singularities of
$f(\mrz)$ arise for values of $\mrz$ for which two singularities of $g(\mrz, s)$ coincide in the $s$ plane,
trapping the contour of integration. A general point on a Landau singularity corresponds to the occurrence 
in the integration space of the analog of this mechanism for functions of several complex variables. The
extension to more than one external variables and the situations where a pinch may become harmless (\eg falling
off the end of the contour) are discussed in chapter~2.1 of \Bref{Eden:1966dnq}; the extension to multiple integrals
can also be found there and in \eqn{mpinch}.

There also  exist  ``second-type''  (so-called non-Landau) singularities (see for example \Bref{Eden:1966dnq}). These arise 
in Feynman loop integrals as pinch singularities at infinite loop momentum and will not be analyzed in this work.

\paragraph{More details on Landau singularities} \hspace{0pt} \\
For more details from the point of view of algebraic geometry see \Brefs{Regge1964,Ponzano:1970ch}.
Landau equations in the context of the theory of asymptotic operation have been discussed in \Bref{Tkachov:1997ap}.
Finally, solutions to the Landau conditions, corresponding to kinematic configurations where the modified Cayley 
determinant vanishes, are called singularities of the first type (singularities of the first type comprise all 
solutions to the Landau conditions for finite values of loop momentum); for a geometric interpretation in terms 
of volumes of polytopes see \Bref{Abreu:2017ptx}. For an interpretation in terms of projective geometry and
momentum twistors see \Bref{Dennen:2015bet}. For the analyticity properties of amplitudes in theories with nonlocal 
vertices of the type occurring in string field theory see \Bref{Chin:2018puw}.
For Landau diagrams in theories with gravity duals see \Bref{Maldacena:2015iua}.
Furthermore, for amplitudes of generalized polylogarithm type there should be a close connection between symbol 
entries and solutions of the Landau equations~\cite{Maldacena:2015iua}. 
The most recent developments deal with massless theories which is of no help here since, in our case,
at least one internal line must have nonzero mass.

Another description of the AT is as follows: $3$ and higher point functions can be cut im more that two pieces; putting all
propagators on-shell corresponds to $\alpha_i \not= 0$ at the level of the Landau equations, \ie ATs go beyond
the concept of unitarity cuts~\cite{Cutkosky:1960sp,Mandelstam:1960zz,RevModPhys.33.448}. 
For an alternative proof of cutting rules in quantum field theories see \Bref{Pius:2018crk}.

Vanishing Cayley determinants have been mentioned in the literature but, usually, this case has not been considered
in detail since ``the exceptional case with a vanishing modified Cayley determinant hardly appears in applications'',
see \Brefs{Denner:2010tr,Fleischer:2011bi}; however, reduction of tensor integrals for small Gram determinant and/or 
small modified Cayley determinant have been discussed in \Bref{Denner:2005nn}.

The expansion of Feynman integrals around their AT is easy to derive analytically 
and only requires Mellin-Barnes and sector decomposition techniques as explained in \Bref{Ferroglia:2002mz}.
Examples of leading behavior (\hyperref[SB]{details are given in~\ref*{SB}}) are:
for the vertex $\mrC_0 \sim \ln\,\Delta_3$; for the box $\mrD_0 \sim \Delta^{-1/2}_4$; 
for the pentagon $\mrE_0 \sim \Delta^{-1}_5$ and no singularity 
for the hexagon $\mrF_0$ in $4\,$dimensions~\cite{Eden:1966dnq}; \eg $\;\Im\,\mrC_0$ has a 
logarithmic singularity, $\;\Re\,\mrC_0$ has a discontinuity. $\Delta_n$ is analytic in the immediate vicinity of the given
singularity and singular on it. Of course, we can add any function, analytic in the neighborhood of the singularity. 
The general result can be summarized as follows: let $\mrL$ be the number of internal lines in the Feynman diagram 
under consideration and $\nu$ the number of loops; define 
$\rho = 2\,\nu - 1/2\,(\mrL + 1)$, the leading behavior of the diagram is given by 
(\hyperref[SB]{special cases are discussed in~\ref*{SB}})
\bq
\Delta^{\rho}_{\mrL} \;\;\mbox{for}\;\; \rho < 0 \spc
\quad 
\Delta^{k+1/2}_{\mrL} \;\;\mbox{for}\;\; \rho = k + \frac{1}{2} \spc
\quad
\Delta^{k}_{\mrL}\,\ln \Delta_{\mrL} \;\;\mbox{for}\;\;\rho = k \spc \quad k \in \Zf^* \spp 
\eq
Therefore for $\mrL = 2\,(2\,\nu + n) - 1$ and $n \in \Zf^{+}$ the AT is a pole of order $n$ for the amplitude, 
\eg a simple pole for the one-loop pentagon, for two-loop diagrams with $9$ propagators \etc In all other cases it is a 
branch point. 

\begin{definition}[One-loop diagrams: summary] Any one-loop diagram is specified by
\bei

\item[\dnuma] $\mrH_{\ssN}$, a $(\mrN - 1)\,\times\,(\mrN - 1)$ matrix whose determinant is the Gram determinant.

\item[\dnumb] The set $\{X_1\,,\,\dots\,,\,X_{\ssN -1}\} = \mathrm{X}_{\ssN}$; we will denote by $\mathrm{X}^{\xord}_{\ssN}$ the
set $\{(X_1,\,\dots\,,X_{\ssN-1}) \in \Rf^{\ssN + 1} \mid 0 < X_{\ssN-1} < \,\dots\, X_1 < 1\}$.

\item[\dnumc] The Bernstein{-}Sato{-}Tkachov~\cite{JB,MS,Tkachov:1996wh} factor $\Delta_{\ssN}$ defined 
in \eqns{Gdef}{BSTf}. The Bernstein theorem~\cite{JB} states that for any polynomial in $\Cf[x]$
there exists a non-zero polynomial $b(s) \in \Cf[s]$ (Berstein{-}Sato polynomial~\cite{JB,MS}) and a differential 
operator $\mrP(s) \in D_n(s)$ such that $\mrP(s)\,\cdot\,f^{s+1} = b(s)\,f^s$. For one-loop diagrams 
$\Delta_{\ssN}$ is the explicit form of $b$, as shown in \Bref{Tkachov:1996wh}.

\item[\dnumd] The set of generalized Mandelstam invariants, $\mathrm{I}$; we will denote by $\mathrm{I}_{\phys}$ a set
of invariants internal to the physical region~\footnote{In this paper ``physical region'' is identified with
the phase space for the corresponding process, \ie the physical region of a given process is the set of all real 
initial and final energy-momenta variables subject to the mass-shell conditions and to energy-momentum conservation. 
Solutions that correspond to points outside the physical region are on the wrong sheet.}. 

\eei

\end{definition}

\begin{definition}[Real (complex) masses: anomalous threshold]
Assume that all masses are real ($\{\Gamma\} = 0$), the physical-region LLS (or physical-region anomalous threshold) 
is given by
\bq
\Delta_{\ssN}\bmid_{\{\Gamma\} = 0} = 0 \spc
\quad
\mathrm{X}_{\ssN} = \mathrm{X}^{\xord}_{\ssN} \spc
\quad
\{\mathrm{I}\} = \{\mathrm{I}\}_{\phys} \spp
\eq
There are cases where the first two conditions are satisfied but Mandelstam invariants are moved to their
complex plane. Nevertheless, their real part can be inside the physical region with a tiny imaginary part;
therefore they can be very close to the boundary.
When internal masses are made complex, \ie $m^2_i \to m^2_i - i\,\Gamma_i\,m_i$, singularities move into the
complex $x\,$-space. We are nevertheless interested in the following configurations:
\[
\begin{array}{ll}
\Re\,\Delta_{\ssN} \approx \Delta_{\ssN}\bmid_{\{\Gamma\} = 0} \approx 0 \spc
\quad & \quad
\Im\,\Delta_{\ssN} \muchless 1 \spc \\
& \\
\{\Re\,X_i\} = \{\Re\,X_i\}^{\xord} \spc
\quad & \quad
\Im\,X_i \muchless 1 \spc 
\end{array}
\]
with $\{\mathrm{I}\} = \{\mathrm{I}\}_{\phys}$.
The introduction of complex masses regularizes the singularity since, in general, $\Im\,\Delta_{\ssN} \not= 0$; 
however another special configuration is possible:
\end{definition}
\begin{definition}[Peierls zeros~\cite{Peierls:1961zz}] \label{dPzero}
They are defined by
\bq
\Re\,\Delta_{\ssN} = \Im\,\Delta_{\ssN} = 0 \spc
\label{Pzero}
\eq
and we look for a set of ``real'' invariants that satisfy \eqn{Pzero}, possibly within the physical region and with
$\{\Re\,X_i\} = \{\Re\,X_i\}^{\xord}$.
The effect of these zeros can be seen by considering a simple example:
\bq
\mrF(y\,,\,\mrz) = \int_{-1}^{+1} dx\,\Bigl[ ( x + i\,y)^2 - \mrz^2 \Bigr]^{-1} \spc
\label{pzfig}
\eq
where $y \in \Rf$ and $\mrz \in \Cf$. We derive
\bqa
\mrF(y\,,\,\mrz) &=& \frac{1}{2\,\mrz}\,\Bigl\{ 
\ln \Bigl[ (\mrz - 1)^2 + y^2 \Bigr] - \ln \Bigl[ (\mrz + 1)^2 + y^2 \Bigr] +
\eta(\mrz -1\,,\,\mrz - 1) \Bigr\} 
\nl
{}&=& \frac{1}{2\,\mrz}\,\Bigl[ \ln \frac{\mrz - 1 - i\,y}{\mrz + 1 - i\,y} -
                                \ln \frac{\mrz + 1 + i\,y}{\mrz - 1 + i\,y} \Bigr] \spc
\label{pzexa}
\eqa
where $\eta$ is the 't Hooft-Veltman eta function~\cite{tHooft:1978jhc} and the $\ln \mrv$ denotes the 
principal branch, $- \pi < \mbox{arg}(\mrv) \le + \pi$. If $\mrz = \alpha + i\,\beta$ with
$-1 \le \alpha \le +1$ and $\beta \ge 0$ we derive
\bq
\mrF(0\,,\,\mrz) \sim \frac{i\,\pi}{\mrz} \spc \qquad \mbox{for} \;\; \mrz \to 0 \spc
\eq
showing a pinch singularity for $\mrz = 0$. When we set $\mrz = 0$ with $y \not= 0$ we obtain
\bq
\mrF(y\,,\,0) = - \frac{2}{1 + y^2} \spc
\eq
without a pinch, \ie one double pole of the integrand at $x = - i\,y$ instead of two simple poles at $x= \pm \mrz$. 
In general, let us consider
\bq
\mrF(\mrz_1\,,\,\mrz_2)= \int_{-1}^{+1} dx\,\Bigl[ (x - \mrz_1)^2 - \mrz^2_2 \Bigr]^{-1} \spc
\eq
where $\mrz_{1,2}$ depend on real external parameters, $s_1\,,\,\dots\,,\,s_n$. Let $\Sigma$ be the hypersurface in
$s\,$-space where $\mrz_2 = 0$; furthermore, let $\Pi_+$ be the hypersurface where $ \Im \mrz_1 = 0$ and
$-1 \le \Re \mrz_1 \le +1$. If we follow a path on $\Pi_+$ and approach $\Sigma$ a pinch will appear; starting with a 
point in $\Sigma$ and not in $\Pi_+$ and following any path on $\Sigma$ will not give a pinch singularity.  

To illustrate the behavior of $\mrF(y\,,\,\mrz)$ we fix $\mrz = 0.9 + i\,10^{-3}$ and scale it with a factor $\lambda$,
showing $\mrF(y\,,\,\lambda\,\mrz)$ as a function of $\lambda$ for $y= 0, 10^{-6}, 10^{-5}$. The result is shown in
\refig{ATfig25}; for $y = 0$ the $1/\mrz$ behavior in the imaginary part is evident. With small $y$ there is no
pole but a discontinuity (a large gap) is present, corresponding to a value of $\lambda$ where the imaginary 
part of the first logarithm in \eqn{pzexa} changes sign.

\begin{figure}[t]
   \centering
   \vspace{-3.cm}
   \includegraphics[width=0.9\textwidth, trim = 30 250 50 80, clip=true]{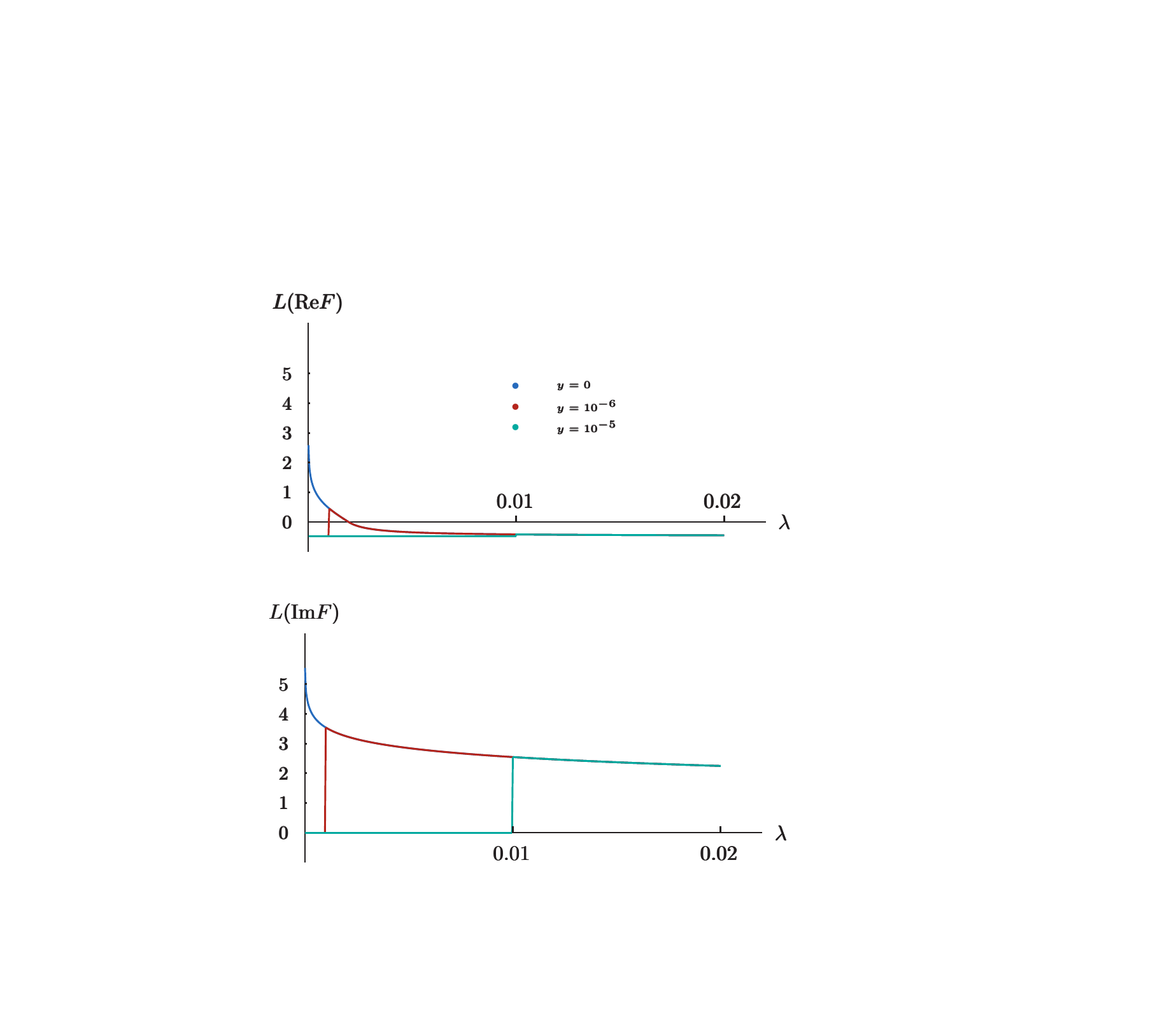}
\vspace{-1.cm}
\caption[]{Real and imaginary part of $\mrL(\mrF)$ where $\mrF$ is given in \eqn{pzfig} and $\mrL$ is the
log-modulus function~\protect\cite{lmod} defined in \eqn{logmod}.}
\label{ATfig25}
\end{figure}

A more detailed discussion concerning boxes and pentagons will be given in \autoref{Scas}.
\end{definition}

\begin{figure}[t]
   \centering
   \vspace{-3.cm}
   \includegraphics[width=1.\textwidth, clip=true]{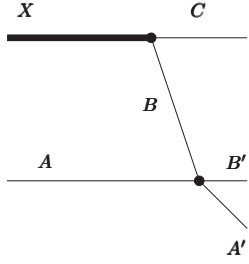}
\vspace{-16.cm}
\caption[]{Diagram which generates the Brayshaw singularity~\protect\cite{Brayshaw:1978xt,Tuan:1978sf}.}
\label{ATfig4}
\end{figure}

\begin{remark}[Analytic continuation]
As long as $\Im\,X_i \not= 0$ there is no singularity, even if $\{\Re\,X_i\} = \{\Re\,X_i\}^{\xord}$.
In general the introduction of complex masses causes the singularity to be removed rather far from the real axis,
\ie the integral 
\bqa
[\mrN]_n(w) &=& \dsimpg{\mrN-1}\,\mrV^{n/2-\mrN}_{\ssN}(\mrw) \spc
\nl
\dsimpg{\mrN-1} &=& \int_{0\,\gamma_1}^1 dx_1\,\int_{0\,\gamma_2}^{x_1} dx_2\,\dots\,
\int_{0\,\gamma_{\ssN-1}}^{x_{\ssN-2}} dx_{\ssN - 1} \spc 
\eqa
is regular if the paths $\gamma_i$, connecting $0$ and $x_{i-1}$ lie on the real axis.
This indicates a branch point of the integral that is not present on the physical sheet but only becomes apparent
after suitable analytic continuation away from the physical contour.
However, there are circumstances where the singularity can be shifted very near (or even inside) the physical region 
defined by $\{\Re\,X_i\} = \{\Re\,X_i\}^{\xord}$, with $\{\Im\,X_i\} = 0$ and 
$\{\mathrm{I}\} = \{\mathrm{I}\}_{\phys}$. In this case we will have the so-called Peierls or Brayshaw 
singularities~\cite{Brayshaw:1978xt,Tuan:1978sf}.
The scattering diagram generating the Brayshaw singularity is shown in \refig{ATfig4} where the blobs refer to 
off-shell scattering amplitudes and $A^\prime, B^\prime$ are at threshold while $X$ is a resonance. 
Consider the $s\,$-plane for the  $X{-}A$ scattering, the Peierls singularity is a special case occurring at 
$s = 2\,(m^2_X + m^2_A) - m^2_B$.
\end{remark}

From \hyperref[CNT]{Coleman-Norton theorem \ref*{CNT}} and \hyperref[KT]{Kershaw theorem \ref*{KT}} we immediately 
realize that a physical-region singularity requires a theory with a 
hierarchy of heavy masses. Furthermore identical masses in a vertex must be avoided, \eg $X$ and $C$ in \refig{ATfig4}
cannot have the same mass; if we stay within the standard model this means that $B$ cannot be neutral, a $\PZ$ or 
a $\PH$ boson. Therefore we are limited to consider only two heavy particles, the $\PQt\,$-quark and the
$\PW\,$-boson. Furthermore, anomalous thresholds in the standard model (SM) prefer the so-called ``off-shell'' region, \eg
$\Pg \Pg$ producing an off-shell Higgs (with a virtuality greater that $2\,m_{\PQt}$) subsequently ``decaying'' into 
four fermions, see \Brefs{Kauer:2012hd,Caola:2013yja}.
The situation changes when we consider BSM models as we will discuss later. 
\subsection{Complex masses \label{Cmass}}
The so-called ``complex-mass scheme'' has been introduced and discussed (in modern times) in 
\Brefs{Denner:2006ic,Actis:2006rb,Actis:2006rc,Denner:2014zga} (for an introduction before the advent of gauge theories 
see \Bref{doi:10.1142/9789812795779_0066}; analiticity in the complex mass shell has been discussed 
in \Bref{Stapp:1976mx} ).
An amplitude with unstable (internal) particles can be regarded as an analytic continuation of the amplitude 
defined by Feynman prescription.

One should remember that unstable states lie in a natural extension of the usual Hilbert space that corresponds to 
the second sheet of the $\mrS\,$-matrix~\footnote{This was the conjecture of Peierls: a pole on the second 
sheet is to be identified with an unstable particle.}; for instance we will have to take the logarithm 
of $\mrz = \mrz_{\mrR} + i\,\mrz_{\mrI}$, where $\mrz$ is the polynomial occurring in the calculation of a given 
Feynman diagram and where, in the limit of zero widths, we have $\mrz = \mrz_0 - i\,0$. The analytic 
continuation requires a new definition~\cite{Passarino:2010qk}, \ie the first Riemann sheet for all quadrants but 
the second where the function is defined in the second Riemann sheet:
\bq
\ln^-(\mrz_{\mrR} + i\,\mrz_{\mrI}\,,\,\mrz_0 - i\,0) = \ln(z_{\mrR} + i\,\mrz_{\mrI}) - 
2\,i\,\pi\,\theta( - \mrz_0)\,\theta( \mrz_{\mrI}) \spp
\eq
It is easily seen that, as far as Feynman diagrams are concerned, $\ln^-(\mrz)$ and $\ln(\mrz)$ coincide when 
the internal masses are made complex while Mandelstam invariants remain real. 

The numerical evaluation of logarithms of complex quantities, when needed, is better performed by using
\bq
\ln x = x\,\mrR_{\mrc}\lpar \frac{1}{4}\,x^2\,,\,x \rpar \spc
\qquad
\arctan x = x\,\mrR_{\mrc}\lpar 1 \,,\, 1 + x^2 \rpar \spc
\eq
where $\mrR_{\mrc}$ is one of the Carlson elliptic integrals~\cite{Carlson}.

In the following sections we will analyze the presence of regularized ATs for bubbles, triangles, boxes \etc and 
study their impact
on physical observables; it should be emphasised that the presence of ATs depends on the structure of denominators
of specific Feynman diagrams~\footnote{Because the vertices are point interactions, singularities in any local QFT
are generated only by propagators.} but their numerical impact on the full amplitude also depends on numerators.
In this regard, we are assuming that the singularity spectrum of the $\mrS\,$-matrix is confined to the 
union of the singularity spectra of the Feynman integrals, and we proceed to construct the singularity 
spectra of the Feynman integrals. Therefore, the scattering amplitude appears as the sum of infinitely many diagrams 
of increasing complexity and each diagram in principle can be completely investigated.
In principle there is no reason for the $\mrS\,$-matrix to be the sum of the diagrams but we work under the
assumption that the diagrams represent the local behavior of the amplitude and that the whole picture can be 
recovered by gluing together all these local behaviors. For an interpretation of the Landau singularities as macroscopic
causality see \Brefs{Chandler:1969bd,Iagolnitzer:1969sk}.
 
Furthermore, at the amplitude level, a given branch cut is generically shared by several integrals 
while the leading singularities receive contributions from a single integral~\footnote{For the most general scenario
this is an assumption, see the discussion on equivalent diagrams in \Brefs{Coster:1969zu,Coster:1970ev}.}.
An additional comment concerns the difference between the Feynman diagram approach and the $\mrS\,$-matrix
theory~\cite{Regge:1977nj}. Feynman integrals can be analytically continued around a Landau singularity: as a 
consequence Feynman diagram integrals are clearly singular only on the Landau surfaces obtained from the 
so-called $+\,\alpha$ Landau equations~\cite{White:2000zs}. In $\mrS\,$-matrix theory there is, a-priori, 
no $i\,\epsilon$ prescription (unless it is adopted as an additional postulate).

It should be emphasized that a complete numerical study of different processes falls outside the scope of this paper
where we limited our analysis to the evaluation of the anomalous part of the amplitudes. Furthermore, we have not
analyzed the impact of parton distribution functions but the general rule is that, given an amplitude (squared),
increasing the number of integrations decreases the effect.
\section{Bubble diagrams \label{bubbles}}
The whole procedure can be understood in terms of a simple example, the generalized bubble integral,
\bq
\mrB_0(\alpha) = \int_0^1 dx \chi^{-\alpha} \spc
\qquad
\chi= s\,x^2 + (m^2_2 - m^2_1 - s)\,x + m^2_1 \spp
\eq
We introduce complex masses, $m^2_i \to m^2_i - i\,\Gamma_i\,m_i$, and derive
\bq
\chi= s\,(x - X)^2 + \Delta_2 \spc
\eq
where we have introduced
\bq
X = \frac{1}{2}\,\lpar 1 + \frac{m^2_1 - m^2_2}{2} \rpar - \frac{i}{2}\,\frac{\Gamma_-}{s} \spc
\quad
\Delta_2 = \lambda(s\,,\,m^2_1\,,\,m^2_2) - \Gamma^2_- +
           2\,i\,\Bigl[ \Gamma_+\,s - \Gamma_-\,(m^2_1 - m^2_2) \Bigr] \spc
\eq
where $\Gamma_{\pm} = \Gamma_1\,m_1 \pm \Gamma_2\,m_2$ and where $\lambda$ is the K\"allen lambda function. 
With real masses the integral is singular when $0 < X < 1$ and $\lambda = 0$, \ie $s = (m_1 \pm m_2)^2$
which are the so-called normal and pseudo threshold.  
We obtain
\bq
\mrB_0(\alpha) = \frac{\Gamma(\alpha - 1/2)}{\Gamma(\alpha)}\,(\frac{\pi}{s})^{1/2}\,\Delta^{1/2 - \alpha}_2 +
\quad \mbox{reg. terms} \spc
\eq
where $\Gamma$ is the Euler gamma function. Furthermore, at the normal threshold (the leading Landau singularity
for the bubble) the condition $0 < X < 1$ is always satisfied. With complex masses there is no singularity
but for
\bq
s = \frac{\Gamma_1\,m_1 - \Gamma_2\,m_2}{\Gamma_1\,m_1 + \Gamma_2\,m^2}\,(m^2_1 - m^2_2) \spc
\eq
we have the Peierls zero ($\Re\,\Delta_2 = \Im\,\Delta_2 = 0$) if
\bq
4\,m^2_1\,m^2_2\,(m^2_1 - m^2_2)\,(\Gamma^2_2 - \Gamma^2_1) = (m^2_1\,\Gamma^2_1 - m^2_2\,\Gamma^2_2)^2 \spp
\label{Bcz}
\eq
To show the numerical effect we select 
$m_1 = 100 \UGeV$, $m_2 = 10 \UGeV$ and $\Gamma_1 = 100 \UMeV$. Deriving $\Gamma_2$ and $s = s_*$ from \eqn{Bcz} we 
find $\sqrt{s_{\ssP\ssT}} - \sqrt{s_*}= 112.5 \UMeV$, where the pseudo-threshold is at $90\,\UGeV$. Numerical results 
are shown in Tab.~\ref{B0tab} where it can be seen that $\Im\,X$ is always negative, $\Re\,X$ decreases with $s$ 
(but $\Re\,X > 1$ at the  pseudo-threshold), there is a cusp at the normal threshold and no special enhancement 
at $\sqrt{s_*}$. 

\begin{table} 
\begin{center}
\resizebox{\columnwidth}{!}{%
\begin{tabular}{lcccccc}
$\sqrt{s} [\UGeV]$ & $\Re \mrB_0(1)$ & $\Im \mrB_0(1)$ & $\Re \Delta_2$ & $\Im \Delta_2$ & $\Re X$ & $\Im X$ \\
\hline
$80$ & $0.78746\,10^{-3}$ & $0.34494\,10^{-5}$ & $-0.59143\,10^{-1}$ & $0.22827\,10^{-3}$ & $1.2734$ & $-0.7031\,10^{-3}$ \\
$\sqrt{s_*}$ & $0.99996\,10^{-3}$ & $0.54997\,10^{-5}$ & $0$ & $0$ & $1.1111$ & $-0.5556\,10^{-3}$ \\ 
$90$ & $0.99996\,10^{-3}$ & $0.54998\,10^{-5}$ & $0.30864\,10^{-6}$ & $-0.16975\,10^{-8}$  & $1.1111$ & $-0.5556\,10^{-3}$ \\
$100$ & $0.15225\,10^{-2}$ & $0.13213\,10^{-4}$ & $0.99752\,10^{-2}$ & $-0.10450\,10^{-3}$ & $0.9950$ & $-0.4500\,10^{-3}$ \\
$110$ & $0.13984\,10^{-1}$ & $0.14968\,10^{-1}$ & $0.13831\,10^{-6}$ & $-0.15026\,10^{-3}$ & $0.9091$ & $-0.3719\,10^{-3}$  \\
$120$ & $-0.72742\,10^{-3}$ & $0.16493\,10^{-2}$ & $-0.17470\,10^{-1}$ & $-0.16710\,10^{-3}$ & $0.8438$ & $-0.3125\,10^{-3}$ \\
\hline
\end{tabular}
}
\end{center}
\caption[]{Peierls zeros for a generalized two-point function. \label{B0tab}}
\end{table}

\section{Triangle diagrams \label{triangles}}
\begin{figure}[t]
   \centering
   \vspace{-3.cm}
   \includegraphics[width=0.9\textwidth, trim = 30 250 50 80, clip=true]{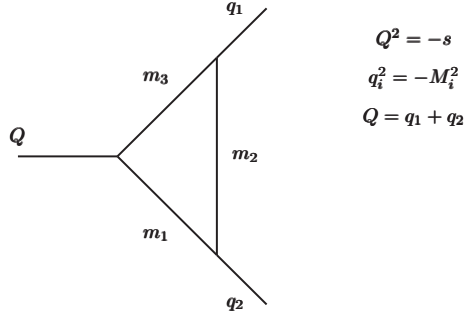}
\vspace{-4.cm}
\caption[]{Triangle diagram: the general case with arbitrary internal and external masses.}
\label{ATfig1}
\end{figure}

Consider the diagram of \refig{ATfig1} where the three external lines are off-shell, \eg $\PH^* \to
\PWs\PWs$. We must have
\bq
s \ge (\mrM_1 + \mrM_2)^2 \spp
\eq

\begin{figure}[t]
   \centering
   \includegraphics[width=1.\textwidth, clip=true]{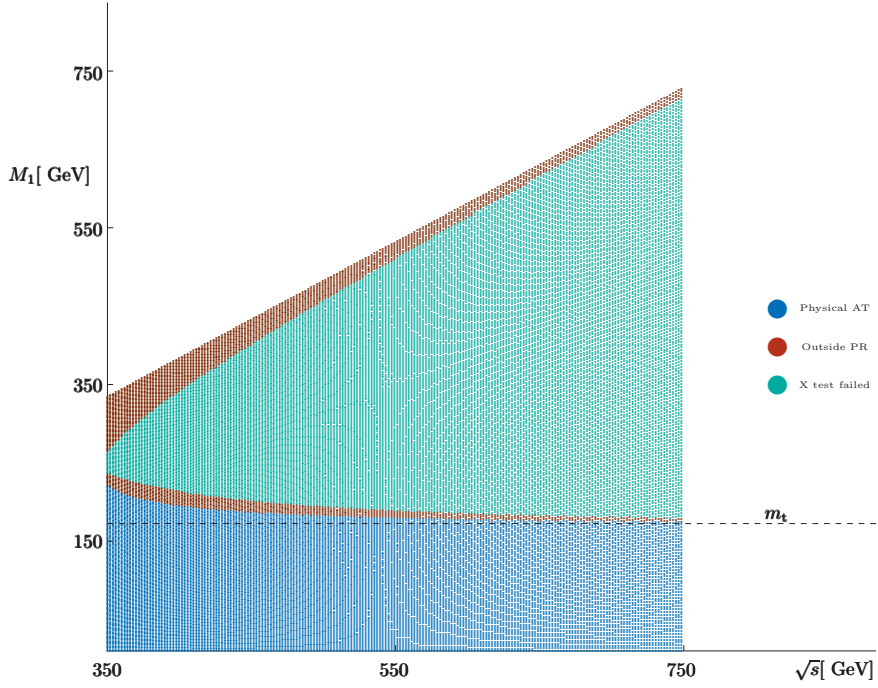}
\vspace{-12.cm}
\caption[]{Physical and non physical AT for the triangle diagram of \refig{ATfig1} with internal masses
given in \eqn{tconf}. Here $350 \UGeV < \sqrt{s} < 750 \UGeV$ and $\mrM_{1,2} > 10 \UGeV$. For given values of
$\sqrt{s}$ and $\mrM_1$ the value of $\mrM_2$ corresponding to the AT is computed.}
\label{ATfig2}
\end{figure}

From the Kershaw theorem~\cite{Kershaw:1971rc} we see that the physical-region Landau curve has six branches 
in the $Q, q_1, q_2$ space, \ie
\bq
s > ( m_1 + m_3)^2 \spc
\quad 
\mrM^2_2 > (m_2 + m_3)^2 \spc
\quad
\mrM^2_1 < (m_1 - m_2)^2 \spc
\eq
with $Q^0 > 0$ and $q^0_2 > 0$. The other branches are obtained by cyclic permutations and by the overall 
reflection of the external momenta. Our example will be as follows: there is an off-shell $\PH$ with
momentum $Q$ going to off-shell $\PW$s; internal lines are $\PQt, \PQb$ quarks, \ie 
\bq
m_1 = m_3 = m_{\PQt} \spc \qquad  m_2 = m_{\PQb} \spp
\label{tconf}
\eq
Furthermore, $\mrM_2 > m_{\PQt} + m_{\PQb}$ and $\mrM_1 < m_{\PQt} - m_{\PQb}$.
In this configuration, when the $\PQt\,$-width is neglected, we obtain $X_{1.2}$, Gram and Cayley determinants,
\bqa
\mrG_3 &=& - \frac{1}{4}\,\mrM^2_2 + \frac{1}{2}\,(s + \mrM^2_1) - \frac{1}{4}\,(s - \mrM^2_1)^2 \spc
\nl
\mrC_3 &=& - \frac{1}{4}\,m^2_{\PQt}\,\mrM^4_2 +
 \frac{1}{4}\,\bigl[ s\,(m^2_{\PQt} + m^2_{\PQb} - \mrM^2_1) + 2\,m^2_{\PQt}\,\mrM^2_1 \bigr]
\nl
{}&-& \frac{1}{4}\,s\,(m^2_{\PQt} - m^2_{\PQb})^2 + \frac{1}{4}\,s\,\mrM^2_1\,(m^2_{\PQt} + m^2_{\PQb}) -
      \frac{1}{4}\,(m^2_{\PQt}\,\mrM^4_1 + m^2_{\PQb}\,s^2) \spc
\nl
\mrG_3\,X_1 &=& \frac{1}{4}\,(s + \mrM^2_1 + m^2_{\PQt} - m^2_{\PQb})\,\mrM^2_2 -
 \frac{1}{4}\,(s - m^2_{\PQt} + m^2_{\PQb})\,s -
 \frac{1}{4}\,(\mrM^2_1 + m^2_{\PQt} - m^2_{\PQb})\,\mrM^2_1 + \frac{1}{2}\,s\,\mrM^2_1 \spc
\nl
\mrG_3\,X_2 &=& \frac{1}{4}\,(\mrM^2_1 + m^2_{\PQt} - m^2_{\PQb})\,\mrM^2_2 +
 \frac{1}{4}\,(\mrM^2_1 - m^2_{\PQt} + m^2_{\PQb})\,s -
 \frac{1}{4}\,(\mrM^2_1 + m^2_{\PQt} - m^2_{\PQb})\,\mrM^2_1 \spc
\eqa
The condition $\Delta_3 = 0$, at $\Gamma_{\PQt} = 0$, can be seen as a quadratic equation in $\mrM^2_2$ for fixed $s$ and 
$\mrM_1$. Therefore, we require
\bqa
{}&[\mrT_1]& \qquad \Delta_3 \,=\, 0 \spc \nl
{}&[\mrT_2]& \qquad 0\, <\, X_2\, < \,X_1\, < \,1 \spc \nl
{}&[\mrT_3]& \qquad s \,>\, (\mrM_1 + \mrM_2)^2 \spp
\label{ATtria}
\eqa
The space-time picture is the following: the state of momentum $Q$ decays into a $\PAQt \PQt$ pair, one
of the top quarks decays into $\PW \PQb$, the $\PQb$ quark rescatters against the second $\PQt$ quark to produce
a state with invariant mass $\mrM_2$. The solution of $\Delta_3 = 0$ can give complex $\mrM^2_2$, $\mrM^2_2$, real 
but negative, a real solution which does not satisfy condition $[\mrT_2]$ in \eqn{ATtria}, a real solution satisfying 
$[\mrT_2]$ (the $X$ test) but not $[\mrT_3]$ (singularity inside the physical region) and, finally, a physical 
singularity satisfying both $[\mrT_2]$ and $[\mrT_3]$.
The distribution of physical-region singularities ($\Gamma_{\PQt} = 0$) is shown in \refig{ATfig2} in the $\sqrt{s}{-}\mrM_1$ 
plane with $350 \UGeV < \sqrt{s} < 750 \UGeV$ and $\mrM_{1,2} > 10 \UGeV$.

To study the corresponding Peierls zeros we introduce $s_{\pm} = \mrM^2_1 \pm \mrM^2_2$ and derive that
\bq
s_+ = \frac{s^2_-}{s} + 2\,( m^2_{\PQt} - m^2_{\PQb} ) \spc
\quad
s^2_- = \frac{s}{2}\,\Bigl\{ s - 4\,m^2_{\PQt} + 8\,m^2_{\PQb} \pm  
\Bigl[ ( s - 4\,m^2_{\PQt} )^2 + 16\,\Gamma^2_{\PQt}\,m^2_{\PQt} \Bigr]^{1/2} \Bigr\}
\eq
satisfy $\Re\,\Delta_3 = \Im\,\Delta_3 = 0$. Neglecting the $\PQb$ width and using the leading-order (LO) value for
$\Gamma_{\PQt}$ we find no Peierls zeros inside the physical region defined by $s > (\mrM_1 + \mrM_2)^2$.
For instance, for $\sqrt{s} = 350\UGeV$, the zero corresponds to invariant masses of $486.3\UGeV$ and
$254.0\UGeV$ or to a negative value for $s^2_-$.

\begin{figure}[t]
   \centering
   \vspace{-3.cm}
   \includegraphics[width=1.\textwidth, clip=true]{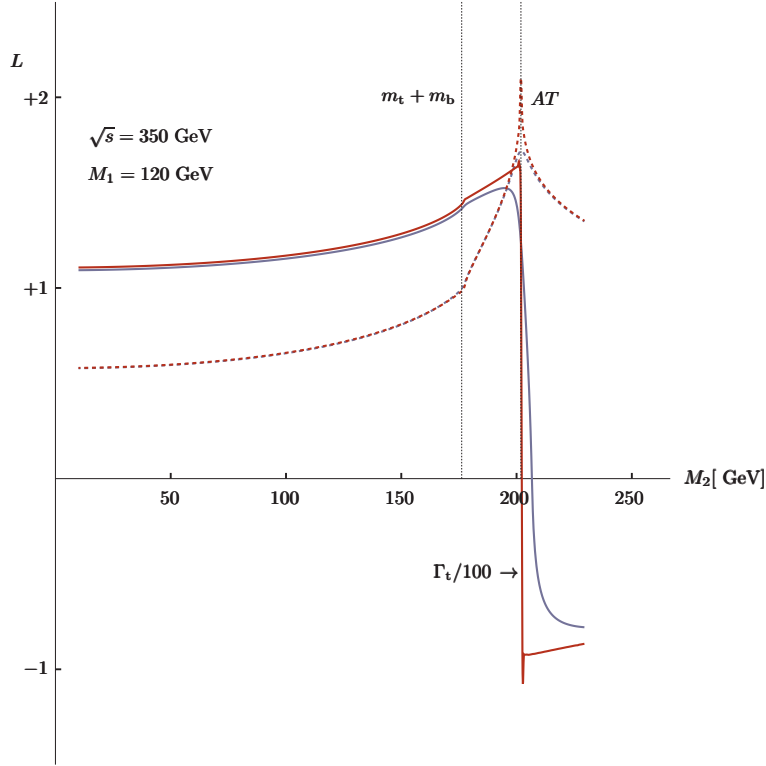}
\vspace{-9.cm}
\caption[]{The log-modulus transformation of $s\,C_0$ as a function of $\mrM_2$ for
$\sqrt{s} = 350\UGeV$, $\mrM_1 = 120\UGeV$. The red curves correspond to $\Gamma_{\PQt}/100$.}
\label{ATfig5}
\end{figure}

We give an example in \refig{ATfig5} where we plot $\mrL(s\,C_0)$ as a function of $\mrM_2$ for 
$\sqrt{s} = 350\UGeV$, $\mrM_1 = 120\UGeV$ and where the log-modulus function~\cite{lmod} is
\bq
\mrL(x)= \mbox{sign}(x)\,\frac{\ln(1 + \mid\,x\,\mid)}{\ln\,10} \spp
\label{logmod}
\eq
The effect of the normal threshold, $\mrM_2 = m_{\PQt} + m_{\PQb}$, and of the anomalous threshold , $\mrM_2 = 201.89\UGeV$
are clearly visible in the real part (solid line) and in the imaginary one (dashed line). Red curves correspond to
the choice of $\Gamma_{\PQt}/100$. 

\begin{figure}[tb]
   \centering
   \vspace{-3.cm}
   \includegraphics[width=1.\textwidth, clip=true]{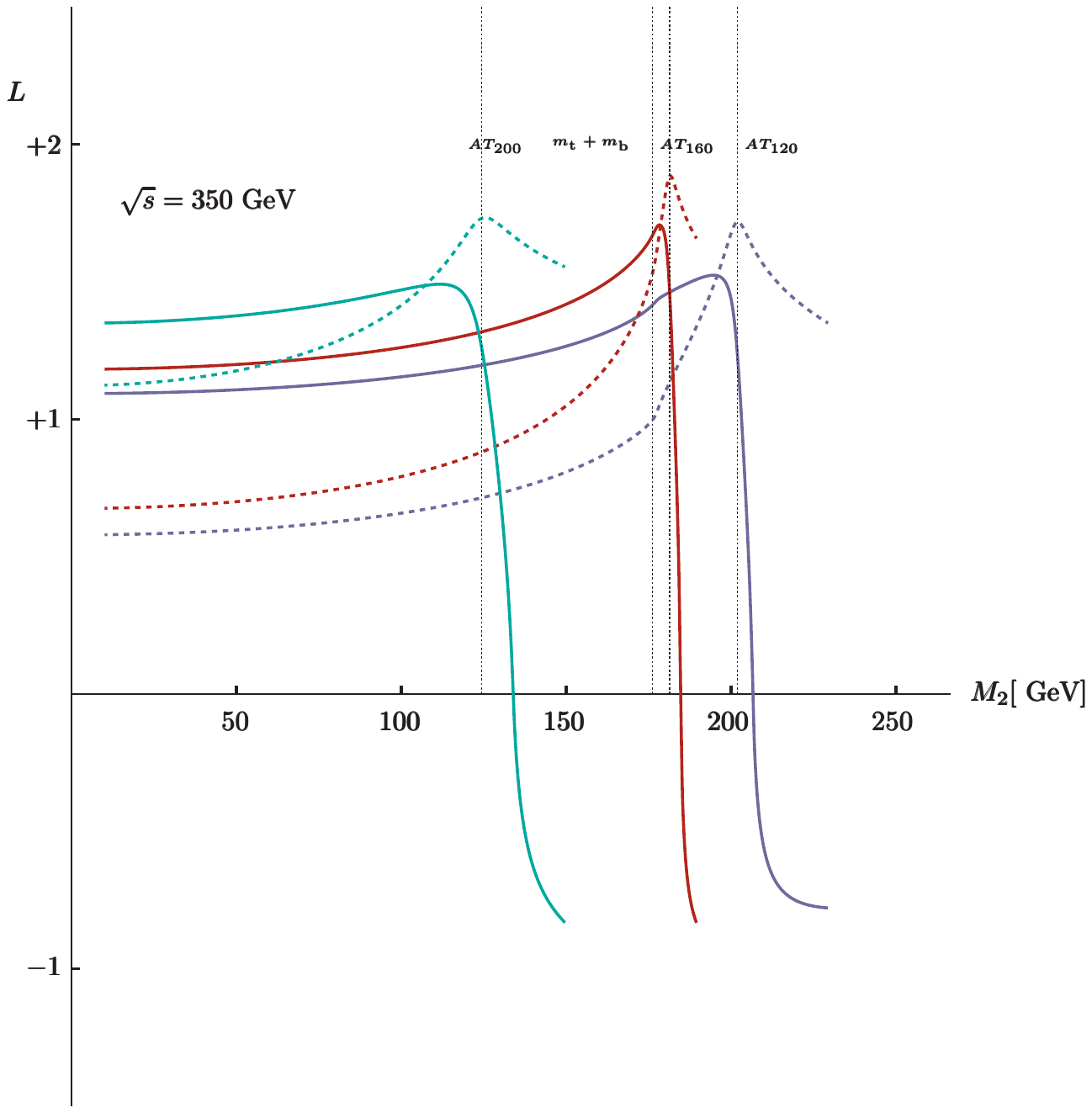}
\vspace{-9.cm}
\caption[]{The log-modulus transformation of $s\,C_0$ as a function of $\mrM_2$ for
$\sqrt{s} = 350\UGeV$, and $\mrM_1 = 120\UGeV$ (blue), $\mrM_1 = 160\UGeV$ (red) and $\mrM_1 = 200\UGeV$ (emerald).
Solid curves give the real part while dashed curves give the imaginary part.}
\label{ATfig6}
\end{figure}

The $\mrM_2$ distribution is also shown in \refig{ATfig6} for $\sqrt{s} = 350\UGeV$ and different values
of $\mrM_1$. It can be seen that higher values for $\mrM_1$ produce lower values for the AT.

\begin{figure}[t]
   \centering
   \vspace{-4.cm}
   \includegraphics[width=1.\textwidth, clip=true]{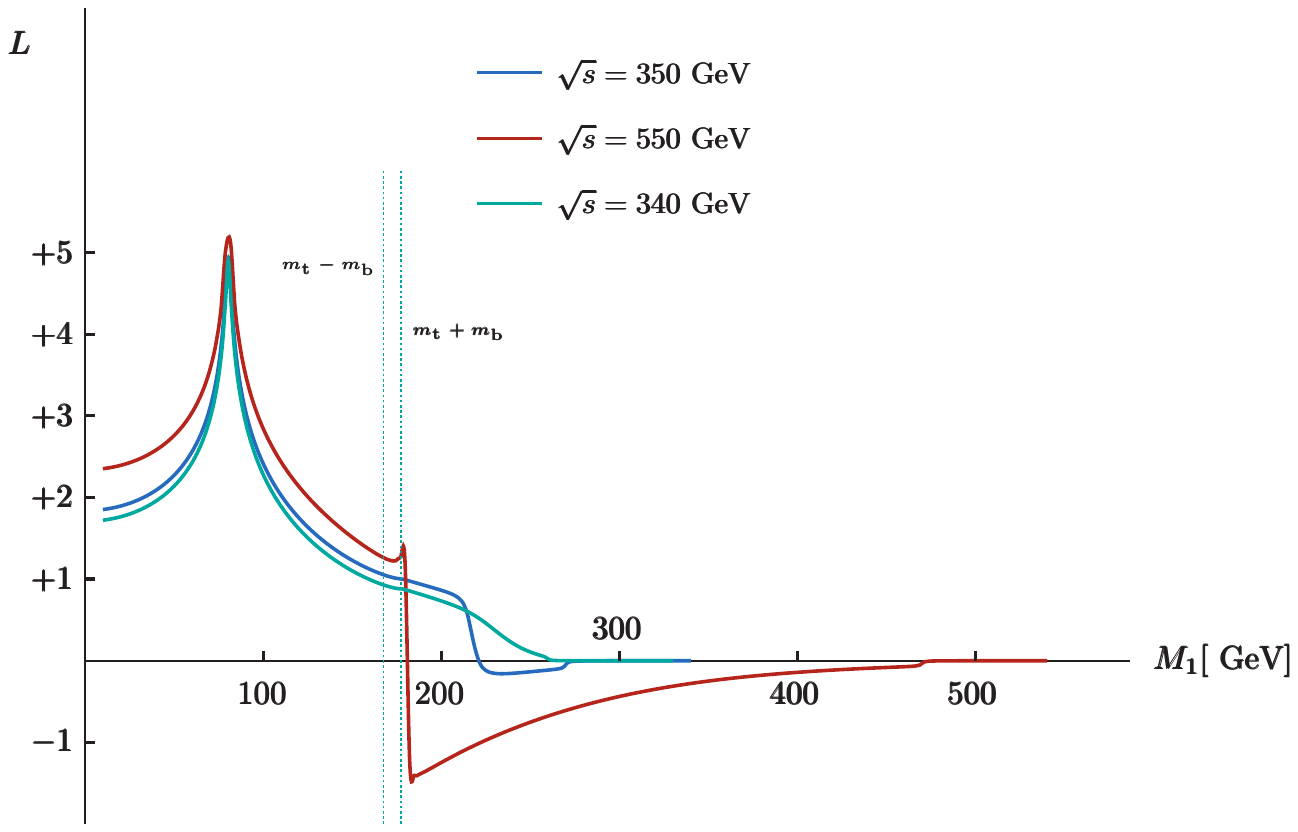}
\vspace{-11.cm}
\caption[]{$\mrL(\mrC_{\Delta})$ as a function of $\mrM_1$ for different values of $\sqrt{s}$. $\mrC_{\Delta}$ is
defined in \eqn{CDdef} and $\mrL(x)$ in \eqn{logmod}.}
\label{ATfig7}
\end{figure}

In order to understand the impact of the AT on some realistic distribution we introduce the $\PW$ complex
pole, $s_{\sPW} = \mu^2_{\sPW} - i\,\mu_{\sPW}\,\gamma_{\sPW}$ and define
\bq
\mrC_{\Delta}(\mrM_1) = s^4\,\Delta_{\sPW}(\mrM_1)\,\int_{\mrM^2_{\mrc}}^{(\sqrt{s} - \mrM_1)^2}\,d\mrM^2_2\,
\Delta_{\sPW}(\mrM_2)\,\Re\,\mrC_0(s\,,\,\mrM^2_1\,,\,\mrM^2_2\,;\,m_{\PQt}\,,\,m_{\PQb}\,,\,m_{\PQt}) \spc
\label{CDdef}
\eq
where the propagator factor is
\bq
\Delta_{\sPW}(\mrM) = \frac{1}{\mid \mrM^2 - s_{\sPW} \mid^2} \spc
\eq
and where $\mrM_{\mrc}$ is a lower cut on $\mrM_2$. We show $\mrL(\mrC_{\Delta})$ (\eqn{logmod}) in \refig{ATfig7} for
$3$ values of $\sqrt{s}$. 

In order to understand the behavior of different curves in \refig{ATfig7} we recall that the occurrence of the
AT requires $\sqrt{s} > 2\,m_{\PQt}$, therefore there is no AT for $\sqrt{s} = 340\UGeV$. Furthermore we must have 
$\mrM_1 > m_{\PQt} + m_{\PQb}$ and $\mrM_2 < m_{\PQt} - m_{\PQb}$ or
$\mrM_1 < m_{\PQt} - m_{\PQb}$ and $\mrM_2 > m_{\PQt} + m_{\PQb}$, with $\mrM_1 + \mrM_2 < \sqrt{s}$. The equation
$\Delta_3 = 0$ is quadratic in $\mrM_2$ and can be solved for fixed $\sqrt{s}, \mrM_1$ and the solutions must
satisfy conditions $\mrT_{2,3}$ in \eqn{ATtria}. We find that at most one of the two solutions does it, as expected.
In particular, for $\sqrt{s}= 350\UGeV$ we always find one real value for $\mrM_2$ that corresponds to the AT if
$\mrM_1 < 168\UGeV$; $\mrM_2$ is complex for $168\UGeV < \mrM_1 < 177\UGeV$ with tiny imaginary part; $\mrM_2$ is
again real for $177\UGeV < \mrM_1 < 226\UGeV$ and becomes imaginary for $\mrM_1 > 226\UGeV$. 

Finally, we consider the process $\Pg \Pg \to \Pem \PGmp \PAGne \PGnGm$, in particular
the $\Pg \Pg \to \PH \to \Pem \PGmp \PAGne \PGnGm$ component. The question of gauge invariance has been
discussed at length in \Bref{Goria:2011wa}: given the process $\Pg \Pg \to \PF$, where $\PF$ is an arbitrary final
state we want to separate the $\PH$ component as
\bq
\sigma_{\Pg \Pg \to \PH \to \PF}(s) = \frac{1}{\pi}\,
\sigma_{\Pg \Pg \to \PH}\,s^2\,\Delta_{\sPH}(s)\,
\frac{\Gamma_{\PH \to \PF}}{\sqrt{s}} \spp
\label{signal}
\eq
where $s_{\sPH}$ is the Higgs complex pole. To summarize: we would like to use the Higgs propagator with its complex pole 
with production and decay computed at arbitrary Higgs virtuality and not at the complex pole.
As far as LO production is concerned, \eg the $\Pg\Pg\PH$ one-loop fermion triangle, there is never an issue of 
gauge-parameter dependence in going off-shell; in this respect higher order QCD corrections are not a problem.

Consider now the decay, \ie $\Gamma(\PH \to \PF)$ in \eqn{signal}: the amplitude $\mrA(\PH \to \PF)$, for each final 
state and as long as we include the complete set of diagrams at one-loop order, is gauge-parameter independent 
if the Higgs boson is on its mass-shell. However, as soon as we put an external leg off-shell, the amplitude
must be coupled to the corresponding physical source and only the complete process $\PI \to \PF$ is gauge-parameter 
independent. The latter does not exclude the existence of subsets of diagrams that satisfy the requirement but 
this can only be examined on a case-by-case basis. To rephrase it, if the Higgs boson is off shell, 
the matrix element still respects gauge invariance (in most cases) in LO and in next-to-leading (NLO) QCD but in 
NLO EW gauge invariance is lost. How to deal with this situation? Technically speaking, we have a matrix element 
\bq
\Gamma\lpar \PH \to \PF \rpar = f\lpar s\,,\,\muhs\rpar,
\eq
where $s$ is the virtuality of the external Higgs boson, $\muh$ is the mass of internal 
Higgs lines and Higgs wave-function renormalization has been included.  
The following happens: $f(s_{\sPH}\,,\,s_{\sPH})$ is gauge-parameter independent 
to all orders while $f(\muhs\,,\,\muhs)$ is gauge-parameter independent at one-loop but not
beyond, $f(s\,,\,\muhs)$ is not. In order to account for the off-shellness of
the Higgs boson we defined a viable scheme by choosing (at one loop level) $f(s\,,\,s)$, \ie
we intuitively replace the on-shell decay of the Higgs boson of mass $\muh$ with the
{\em on-shell} decay of an Higgs boson of mass $\sqrt{s}$ and not with the off-shell decay
of an Higgs boson of mass $\muh$. The same applies for the NLO EW correction to production.

In our case we are interested in the effect of the AT present in the triangle graph $\PH \PWW$ with internal
quark lines and, therefore, there is no problem in the off-shellness of the process (there are no internal Higgs lines); 
furthermore, no other diagram produces an AT (at the same location). As a consequence we can analyze
$\PH \to \Pem \PGmp \PAGne \PGnGm$ with off-shell $\PH$ and look for the impact of the AT on
$\Gamma(\PH \to \Pem \PGmp \PAGne \PGnGm)$.
The process is now
\bq
\PH(Q) \to \PWp(q_1 + q_2) + \PWm(q_3 + q_4) \to \PGnGm(q_1) + \PGmp(q_2) + \Pem(q_3) + \PAGne(q_4) \spc
\label{triaproc}
\eq
with $Q^2= - s$ and light fermion masses are neglected. The full process is given in terms of $s$ and $5$
additional Mandelstam invariants~\cite{Kumar:1970cr},
\bqa
s_1 &=& - (q_2 + q_3 + q_4)^2 \spc \qquad
s_2 = - (q_3 + q_4)^2 \spc
\nl
u_1 &=& - (q_1 + q_3 + q_4)^2 \spc \qquad
u_2 = - (q_1 + q_2 + q_4)^2 \spc
\nl
t_2 &=& - (q_1 + q_4)^2 \spp
\eqa
Next we define $s_2 = \mrM^2$ and study the $d\Gamma/d\mrM$ distribution,
\bq
\frac{d \Gamma}{d \mrM} = \frac{d \Gamma}{d \mrM}\bmid_{\myLO} + \frac{d \Gamma}{d \mrM}\bmid_{\myNLO\,,\,\AT} +
\frac{d \Gamma}{d \mrM}\bmid_{\myNLO\,,\,\rest} \spp
\label{pATdef}
\eq
Since our interest is, in the first place, on the AT effect we limit the calculation to $\delta_{\AT}$, the
percentage correction introduced by the AT. 

\begin{figure}[t]
   \centering
   \vspace{-4.cm}
   \includegraphics[width=1.\textwidth, clip=true]{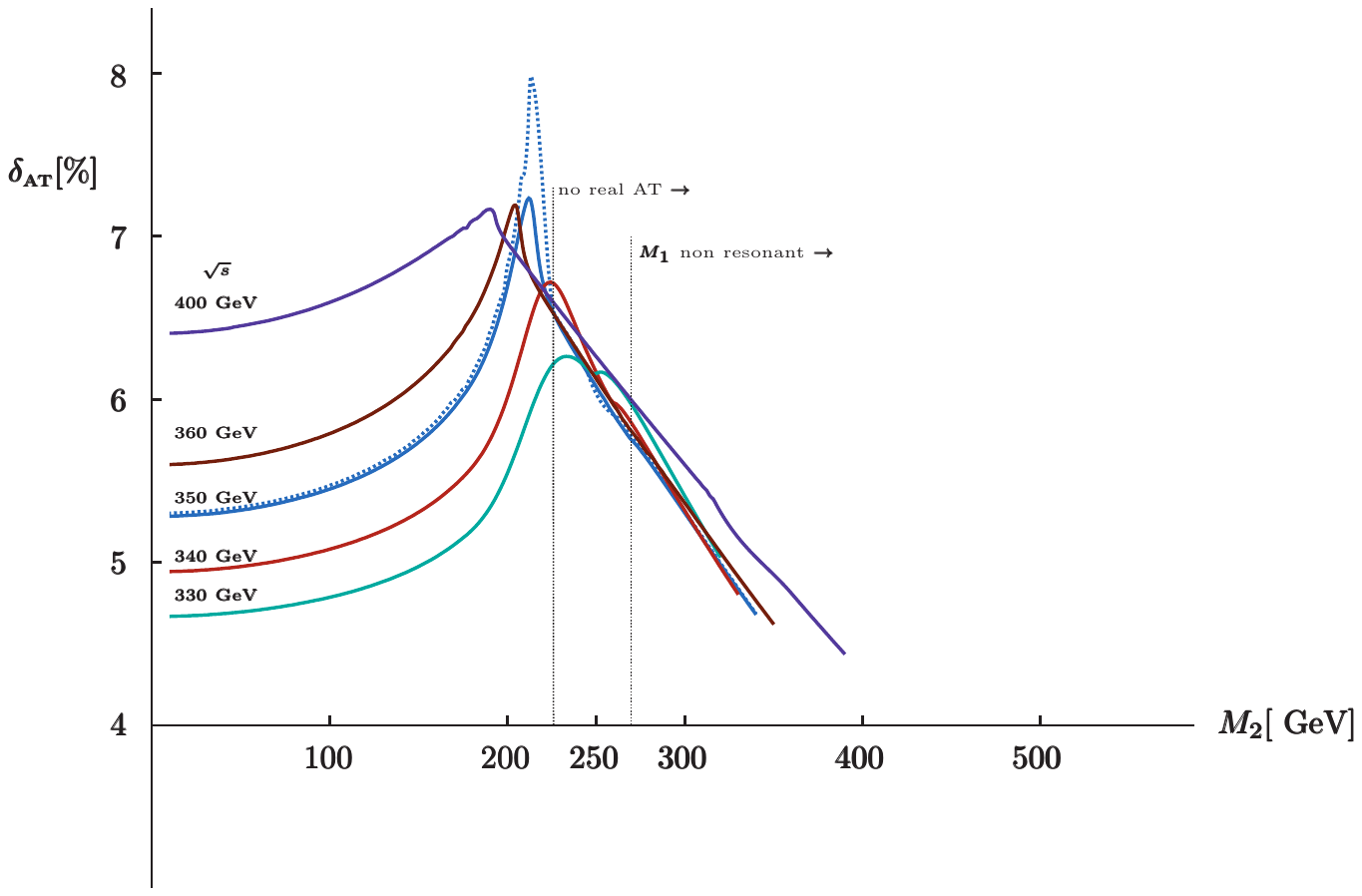}
\vspace{-11.cm}
\caption[]{Percentage radiative corrections around the AT (\eqn{pATdef}) for the process of \eqn{triaproc}.}
\label{ATfig11}
\end{figure}

Results are shown in \refig{ATfig11} for different values of $\sqrt{s}$. To understand the behavior of radiative
corrections two effects should be taken into account: $\mrM_1$ is non resonant if $\mrM_2 > \sqrt{s} - \mw$ and
above a certain value for $\mrM_2$ there is no AT corresponding to a real value of $\mrM_1$. For instance, for
$\sqrt{s} = 350\UGeV$ this happens above $\mrM_2 \approx 226\UGeV$. The (blue) dashed line corresponds to
$\sqrt{s} = 350\UGeV$ and $\Gamma_{\PQt} \to \Gamma_{\PQt}/100$, showing the effect of the AT.
\subsection{AT induced by QCD/QED radiation \label{nsing}}
\begin{figure}[tb]
   \centering
   \vspace{-4.cm}
   \includegraphics[width=1.\textwidth, clip=true]{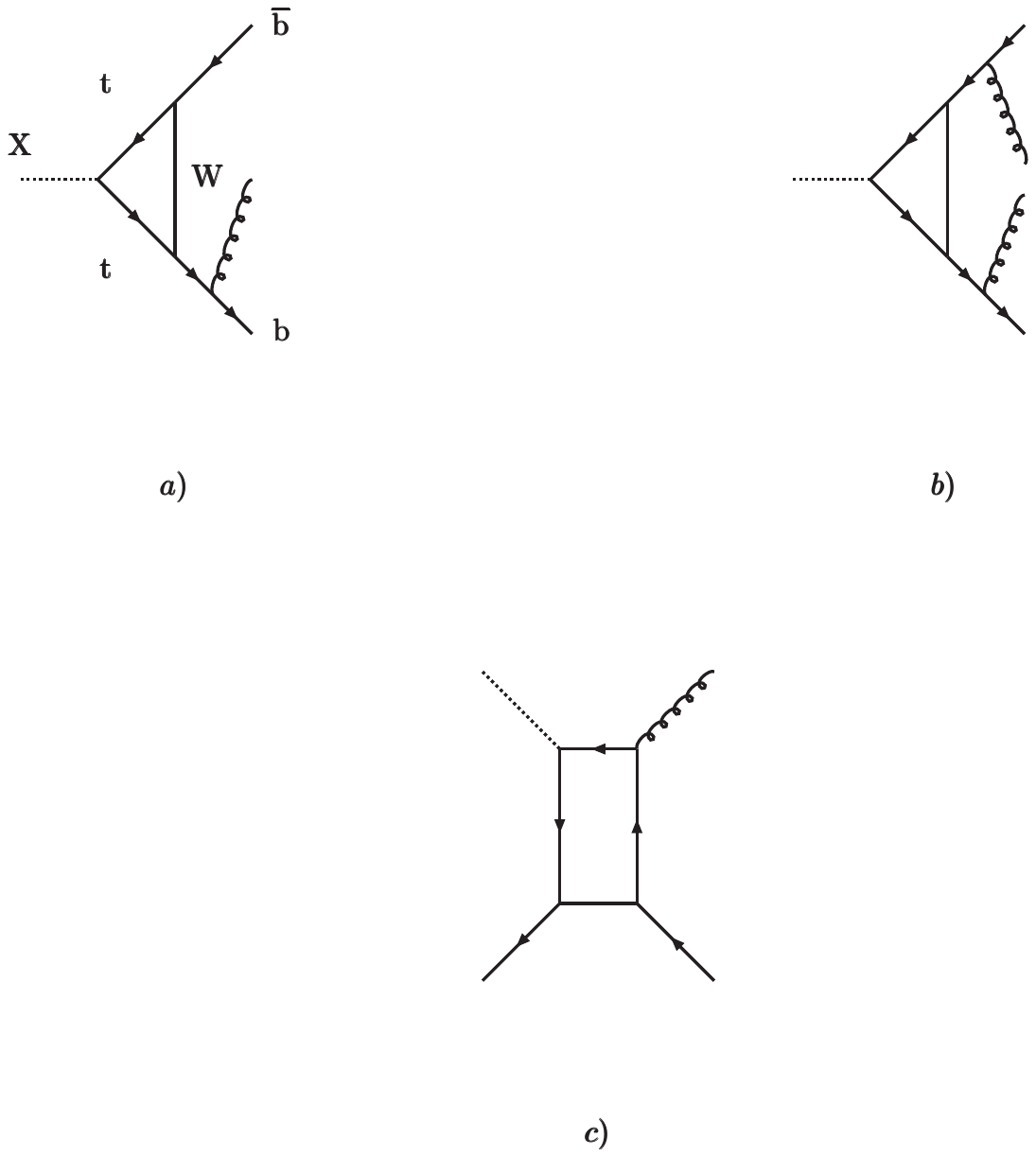}
\vspace{-11.cm}
\caption[]{Example of diagrams contributing to $\PX \to \PAQb \PQb \Pg$ and $\PX \to \PAQb \PQb \Pg \Pg$.}
\label{ATfig22}
\end{figure}

It is easy to see that there is no AT for $\PX \to \PAf \Pf$, where $\PX$ can be an on-shell BSM Higgs boson,
the off-shell SM Higgs boson or an off-shell $\PZ$ boson. However, considering the processes
\bq
\PX \to \PAQb \PQb \Pg \spc
\qquad
\PX \to \PAf \Pf \PGg \spc
\label{Dalproc}
\eq
we find a class of one-loop diagrams (a representative is a) in \refig{ATfig22})
which admit a physical-region AT. The processes shown in \eqn{Dalproc} present several important features,
as discussed in \Brefs{Passarino:1987za,Passarino:2013nka}: if $\PX$ is a Higgs boson the tree-level coupling
$\PH{-}\PAf{-}\Pf$ is Yukawa suppressed, \ie proportional to $m_{\Pf}$. This property is preserved in higher loops; 
however this is not the case when a photon (gluon) is emitted and, already at one loop, there are contributions
surviving the $m_{\Pf} \to 0$ limit. We assume that the BSM (heavy) Higgs boson has couplings proportional to
the SM ones; for instance, in the singlet extension of the SM the $\PH \PW \PW$, $\PH \upphi \upphi$ and
$\PH \PAf \Pf$ couplings are equal to the corresponding SM couplings times $\sin \alpha$ where $\alpha$ is
the mixing angle between the (SM) doublet and the singlet.
\paragraph{ \texorpdfstring{$\PAQb \PQb \Pg(\Pg)$}{} final state} \hspace{0pt} \\
If $s$ denotes the $\PX$ virtuality, we find a small window between 
$\sqrt{s} = 345.5\UGeV$ where the AT corresponds to $\mrM(\PQb \Pg) = 264.81\UGeV$ and
$\sqrt{s}= 370\UGeV$ where the AT corresponds to $\mrM(\PQb \Pg) = 252.90\UGeV$. 
Diagram c) in \refig{ATfig22} is the representative of a class not supporting a (physical-region) LLS,
\ie the leading singularity of diagram a) in \refig{ATfig22} is only subleading for diagram c), corresponding to
the contraction of an internal top line.

\begin{figure}[tb]
   \centering
   \vspace{-4.cm}
   \includegraphics[width=1.\textwidth, clip=true]{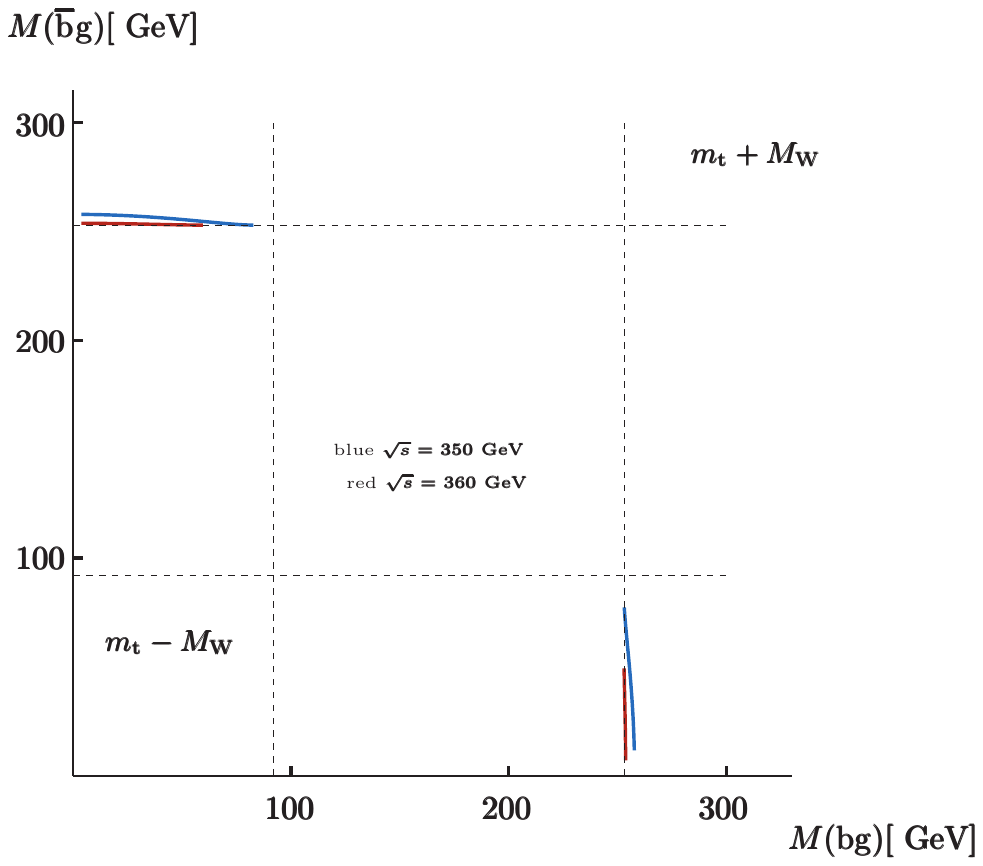}
\vspace{-11.cm}
\caption[]{Anomalous threshold in $\PX \to \PAQb \PQb \Pg \Pg$.}
\label{ATfig23}
\end{figure}

Consider now the process $\PX \to \PAQb \PQb \Pg \Pg$, the double gluon emission (\eg diagram b) in \refig{ATfig22}).
Also in this case there is a physical-region AT as illustrated in \refig{ATfig23} where we show $\mrM(\PAQb \Pg)$ as
a function of $\mrM(\PQb \Pg)$ at the AT; as expected, one of the invariant masses should be above 
$m_{\PQt} + \mrM_{\sPW}$ with the other below $m_{\PQt} - \mrM_{\sPW}$. It is worth noting that even in this case there is only
a small window above $\sqrt{s} = 2\,m_{\PQt}$ where the AT shows up, as illustrated by the red curve in \refig{ATfig23}
which corresponds to $360\UGeV$. The physical-region AT disappears around $\sqrt{s} = 370\UGeV$.

However, above $370\UGeV$ Peierls zeros start to appear, \eg at $\sqrt{s} = 370\UGeV$ the zero corresponds to
one invariant mass of $110.36\UGeV$ with the other at $200.25\UGeV$. At $\sqrt{s} = 512\UGeV$ the values are
$9.73\UGeV$ and $245.63\UGeV$; above this value the two invariant masses are outside the physical region.

\begin{figure}[tb]
   \centering
   \includegraphics[width=1.\textwidth, clip=true]{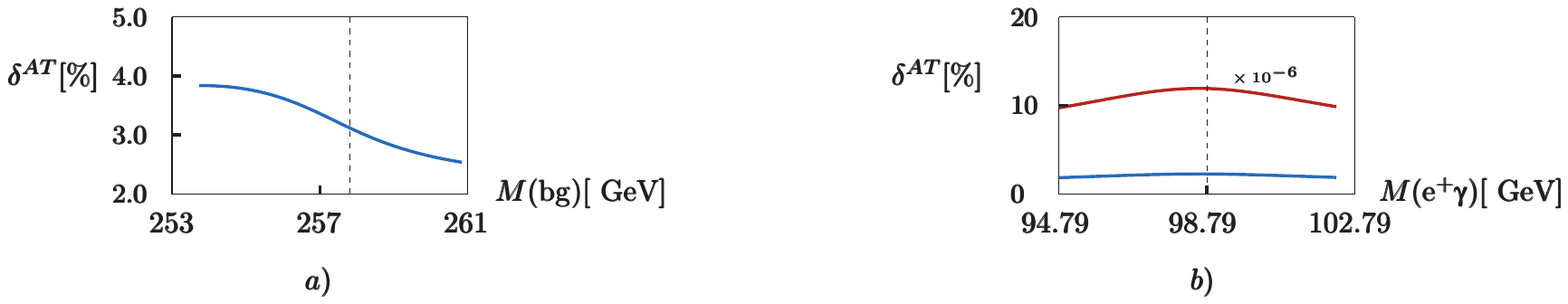}
\vspace{-18.cm}
\caption[]{AT induced radiative corrections for a) $\PH^* \to \PAQb \PQb \Pg$ and b) $\PH^* \to \Pep \Pem \PGg$. 
The $\PH$ virtuality is a) $\sqrt{s} = 350\UGeV$ and b) $\sqrt{s} = 170\UGeV$.}
\label{ATfig24}
\end{figure}

The same line of argument applies to other processes, \eg $\Pg \Pg \to \PAQb \PQb \Pg$ and 
$\Pg \Pg \to \PAQb \PQb \Pg \Pg$ (also to a $\PAQq \PQq$ initial state).
Let us consider $\PH^* \to \PAQb \PQb \Pg$ with an off-shell Higgs boson: we find the following pattern for the
amplitude,
\bq
\mrA_{\myLO} = \mrA^{(1)}_{\myLO}\,m_{\PQb} + \ords{m^2_{\PQb}} \spc
\qquad
\mrA_{\myNLO} = \mrA^{(0)}_{\myNLO} + \mrA^{(1)}_{\myNLO}\,m_{\PQb} + \ords{m^2_{\PQb}} \spp
\eq
Therefore, $\mid \mrA \mid^2$ is $\ords{m^2_{\PQb}}$ at LO; there is no interference between
$\mrA^{(1)}_{\myLO}$ and $\mrA^{(0)}_{\myNLO}$ so that the interference (NLO) is also $\ords{m^2_{\PQb}}$.
Since $\mrA_{\myNLO}$ contains $\mrA^{\AT}_{\myNLO}$ we can include an additional (NNLO) term given by
the square of $\mrA^{(0)}_{\myNLO}$. Taking into account the logarithmic nature of the AT, the suppression factor
$1/(16\,\pi^2)$ for the loop and the value of the $\PQb\,$-quark mass we do not find any significant effect due
to the AT. The process $\PH^*(P) \to \PQb(q_1) + \PAQb(q_2) + \Pg(q_3)$, with $P^2= - s$ is described by
two invariants,
\bq
s_1 = - (P - q_1)^2 \spc
\qquad
u_1 = - (P - q_2)^2 \spc
\eq
with the following boundaries
\bq
m^2_{\PQb} \le s_1 \le (\sqrt{s} - m_{\PQb})^2 \spc
\qquad
u_{1-} \le u_1 \le u_{1+} \spc
\eq
where the limits for $u_1$ are
\bq
u_{1\pm} = s + m^2_{\PQb} - \frac{1}{2\,s_1}\,(s_1 + m^2_{\PQb})\,(s + s_1 - m^2_{\PQb})
\pm \frac{1}{2\,s_1}\,\lambda^{1/2}(s_1\,,\,m^2_{\PQb}\,,\,0)\,
\lambda^{1/2}(s\,,\,s_1\,,\,m^2_{\PQb}) \spp
\eq
After inserting the relevant parts of the one-loop corrections we obtain the percentage AT corrections to
the pseudo-observable $d\Gamma(\PH^* \to \PAQb \PQb \Pg)/d\mrM(\PQb \Pg)$, \ie
$\delta^{\AT}_{\myNLO\,,\,\myNNLO}$. For $\sqrt{s} = 350\UGeV$ the AT is at $\sqrt{s_1} = \mrM(\PQb \Pg) 
\approx 257.82\UGeV$. After imposing a cut of $10\UGeV$ on all final state invariant masses we obtain
the result shown in the left panel of \refig{ATfig24}; NLO and NNLO are indistinguishable. The AT effect is very small 
and corresponds to a change in the curvature of $\delta^{\AT}$.  

\paragraph{ \texorpdfstring{$\Pep \Pem \PGg$}{} final state} \hspace{0pt} \\
For this process we can have a $\PW{-}\PGn{-}\PW$ triangle: in this case $\sqrt{s}$ must be above $2\,\mw$ for a 
physical-region AT, as dictated by the Coleman-Norton theorem. 
The physical-region AT starts at $\sqrt{s} = 161.05\UGeV$ with $\mrM(\Pep \PGg) = 110.64\UGeV$ and approaches
$\mrM(\Pep \PGg) = \mw$ for very large values of $\sqrt{s}$. It is worth noting that all diagrams contributing to
$\PH^* \to \PZ ( \to \Pep\Pem) + \PGg$ do not show an AT. For this process, due to the small value of $m_{\Pe}$ the
inclusion of NNLO terms changes drastically the result as shown in the right panel of \refig{ATfig24}, \ie the
NNLO contribution is six orders of magnitude larger than the LO term at $\sqrt{s} = 170\UGeV$. Here the red curve 
corresponds to no cut while the red one corresponds to a cut of $10\UGeV$ on all final state invariant masses.
In this case $\delta^{\AT}$ shows a peak at the AT.

We can also have a $\PZ{-}\Pe{-}\PZ$ triangle: in this case $\sqrt{s}$ must be above $2\,\mz$. The physical-region AT starts
at $\sqrt{s}= 182.38\UGeV$ with $\mrM(\Pep \PGg) = 128.96\UGeV$ and approaches $\mrM(\Pep \PGg) = \mz$ for very
large values of $\sqrt{s}$. 

\paragraph{ \texorpdfstring{$\Pep \Pem$}{} colliders} \hspace{0pt} \\
It is worth noting that the same qualitative behavior will be found for
\bq
\Pep + \Pem \to \PZ/\PGg \to \PAQb + \PQb + \Pg \spc
\qquad
\Pep + \Pem \to \PZ/\PGg \to \PAf + \Pf + \PGg \spc
\eq
and also for $\Pep + \Pem \to \PAQb + \PQb + \PGg + \PGg$, an irreducible background process in measuring
the $\PH \to \PGg \PGg$ decay at a linear collider.
\section{Box diagrams \label{boxes}}
\begin{figure}[t]
   \centering
   \vspace{-2.cm}
   \includegraphics[width=1.\textwidth, clip=true]{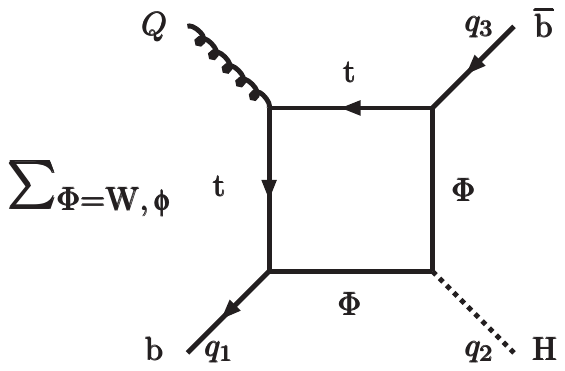}
\vspace{-16.cm}
\caption[]{Family a) for $\Pg(Q) \to \PQb(q_1) + \PAQb(q_3) + \PH(q_2)$.}
\label{ATfig3}
\end{figure}

Consider the diagram of \refig{ATfig3} where $Q = q_1 + q_2 + q_3$, with $Q^2 = - s$ and $q^2_i = - \mrM^2_i$.
The physical region is defined in terms of invariants~\cite{Kumar:1970cr},
\bq
s_1 = - (Q - q_1)^2 \spc \quad u_1 = - (Q - q_2)^2 \spp
\label{boxmi}
\eq
We derive
\bq
(\mrM_2 + \mrM_3)^2 \le s_1 \le (\sqrt{s} - \mrM_1)^2 \spc \qquad
u_{1-} \le u_1 \le u_{1+} \spc
\label{DPR}
\eq
where the limits for $u_1$ are
\bq
u_{1\pm} = s + \mrM^2_2 - \frac{1}{2\,s_1}\,\Bigl[
(s_1 + \mrM^2_2 - \mrM^2_3)\,(s + s_1 - \mrM^2_1) \mp \lambda^{1/2}(s_1\,,\,\mrM^2_2\,,\,\mrM^2_3)\,
\lambda^{1/2}(s\,,\,s_1\,,\,\mrM^2_1) \Bigr] \spp
\label{u1lims}
\eq
Therefore, we are interested in the process
$\Pg \Pg \to \PAQb \PQb \PH$ where $\PH$ can be an off-shell Higgs boson of the SM or some,
on-shell, heavy Higgs boson present in some BSM model. Another case of interest is represented by
$\Pg \Pg \to \PAQb \PQb \PH \PH$; here the two SM Higgs bosons are on shell, therefore $q_2 = q_{21} + q_{22}$
with $q^2_{2i} = - \mrM^2_{\sPH}$ and $\mrM_2$ is the invariant mass of the $\PH \PH$ pair.
Using $\mrM_{1,3} = m_{\PQb}$ and $\mrM = \mrM_2$ we compute the corresponding $\Delta_4$ using $\mrL_4 = m^2_{\PQt}$
and
\bq
\mrG_4 = - \frac{1}{4}\,\lpar a\,\mrM^4 + b\,\mrM^2 + c \rpar \spc
\eq
\bq
a = m^2_{\PQb} \spc \quad
b = (s - s_1)\,u_1 - (2\,s + u_1)\,m^2_{\PQb} \spc \quad
c = (s_1 + u_1 - s)\,s_1\,u_1 + (s^2 - s\,u_1 - 2\,s_1\,u_1)\,m^2_{\PQb} + u_1\,m^4_{\PQb} \spp
\eq

\begin{figure}[t]
   \centering
   \includegraphics[width=1.\textwidth, clip=true]{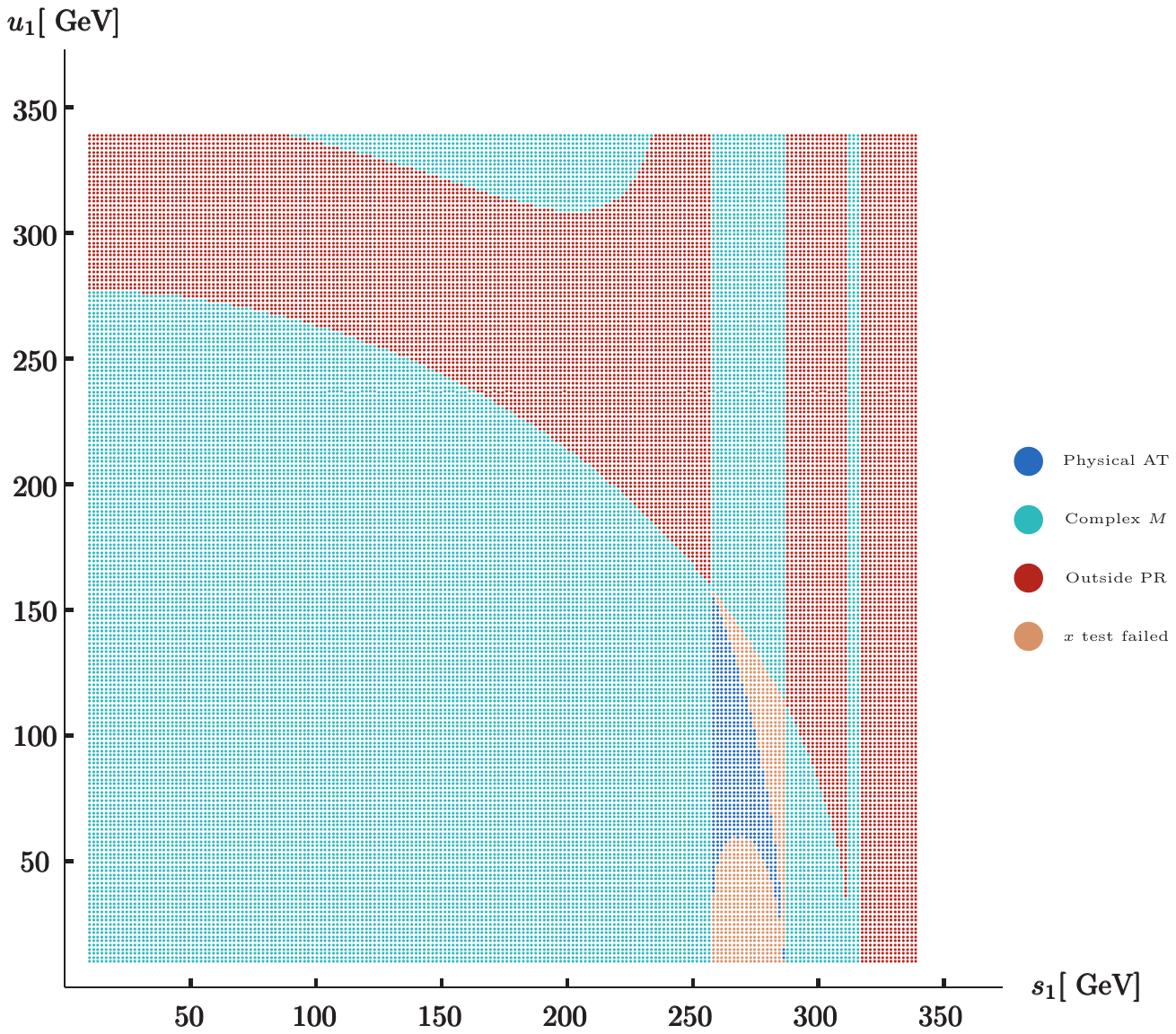}
\vspace{-12.cm}
\caption[]{A scan in the $s_1{-}u_1$ plane searching for a physical-region AT in $\Pg \Pg \to \PAQb \PQb \PH (\PH \PH)$.}
\label{ATfig10}
\end{figure}

It can be seen that $\Delta_4 = 0$ produces a quadratic equation in $\mrM^2$ that can be solved for fixed
$s, s_1, u_1$. A scan in the $s_1{-}u_1$ plane is shown in \refig{ATfig10} for $\sqrt{s} = 350\UGeV$;
the blue region shows values for $\mrM^2$ which correspond to a physical-region AT, \ie $\Delta_4 = 0$, ordered values for
$X_i$ ($0 < X_3 < X_2 < X_1 < 1$) and a real value for $\mrM$ inside the physical region given in \eqn{DPR}.
 

Next we introduce complex poles for $\PW$ and $\PQt$ and look for Peierls zeros. The two equations, $\Re\,\Delta_4 = 0$
and $\Im\,\Delta_4 = 0$ are quadratic in $u_1$ and $\mrM^2$ for fixed $s$ and $s_1$ generating four solutions. To
give an example we fix $\sqrt{s} = 350\UGeV$ and derive
\begin{itemize}

\item[{\underline{$\Gamma = 0$}}] one of the solutions for a physical-region AT is 
$\sqrt{s_1} = 259.15\UGeV$, $\sqrt{u_1} = 38.05\UGeV$ and $\mrM= 162.46\UGeV$ 

\item[{\underline{$\Gamma \not= 0$}}] for $\sqrt{s_1} = 259.15\UGeV$, one solution returns a negative value 
for $u_1$ and the other three (two are coincident) return $\mrM= 172.87\UGeV, \sqrt{u_1} = 58.13\UGeV$ and
$\mrM= 191.84\UGeV, \sqrt{u_1} = 158.16\UGeV$, both outside the physical region.

\end{itemize}

\begin{figure}[t]
   \centering
   \includegraphics[width=1.\textwidth, clip=true]{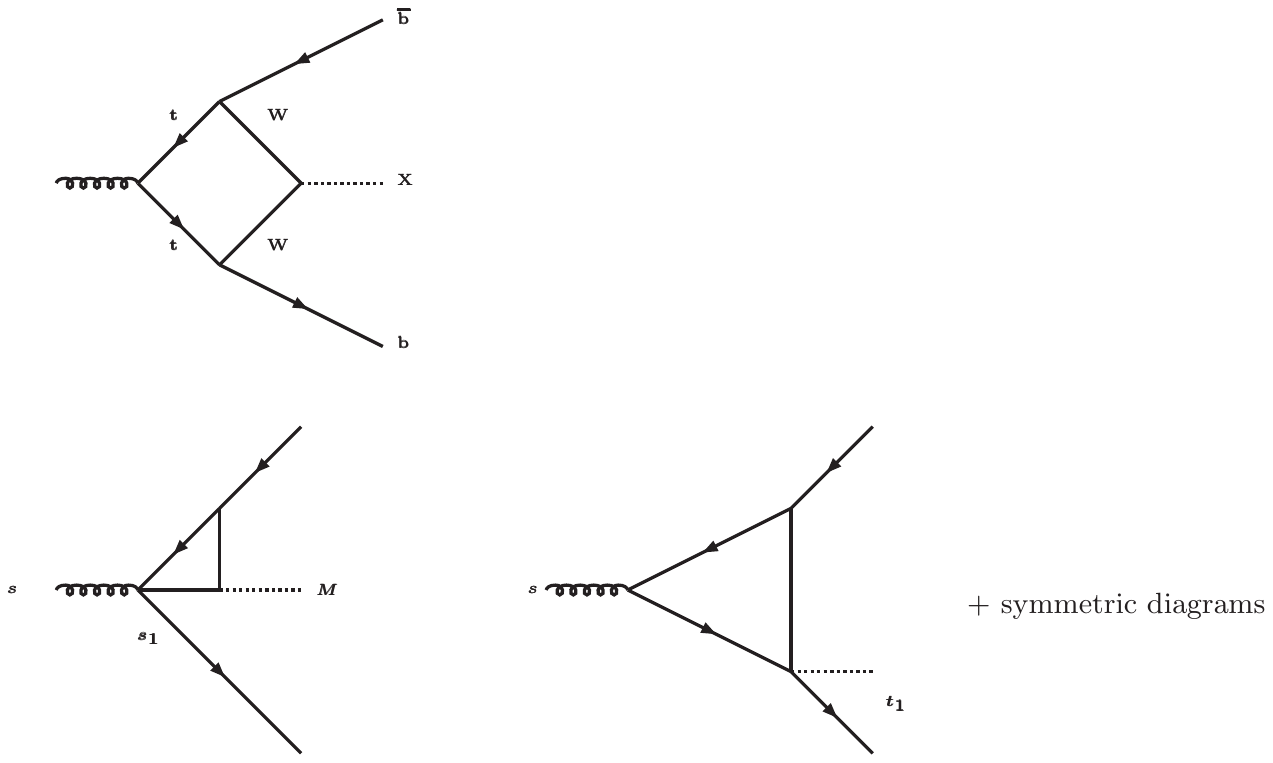}
\vspace{-15.cm}
\caption[]{Box diagram producing an AT for $\Pg \to \PAQb \PQb \PX$ where $\PX$ can be some heavy BSM Higgs boson,
an off-shell SM Higgs boson or a pair of two (on-shell) SM Higgs boson. In the second part of the figure we show two
of the four triangles obtained by shrinking one line of the box to a point.}
\label{ATfig12}
\end{figure}

Any box diagram is decomposed into a box in $6$ dimensions and four triangles: an example is given in 
\refig{ATfig12}. As a consequence we have to look not only for the LLS of the box
but also for the subleading ones, which are the leading singularities for the triangles. 

Old examples can be found in \Bref{PhysRev.160.1346} for the process $\PAp \Pp \to \PK \PK \PGp \PGp$
where peaks were predicted for the amplitude squared in a certain range of the external variables.

We start our analysis by looking at $\Pg(Q) \to \PQb(q_1) + \PAQb(q_3) + \PX(q_2)$ at the point
\bq
\mrP = \{\sqrt{s} = 350\UGeV\,\,;\,\,\sqrt{s_1} = 259.15\UGeV\,\,;\,\,\sqrt{u_1} = 155.20\UGeV\} \spc
\label{refP}
\eq
as a function of $\mrM^2 = - q^2_2$. It is worth noting that the location of the AT does not depend on the
nature of $\PX$, but its numerical impact on the amplitude does. 

For zero widths there is a physical-region AT at $\mrM = 187.72\UGeV$; the Peierls zeros corresponding to 
$\sqrt{s_1} = 259.15\UGeV$ are located at $\sqrt{u_1} = 58.13\UGeV, \mrM = 172.87\UGeV$ and 
$\sqrt{u_1} = 158.16\UGeV, \mrM = 191.84\UGeV$. They are both outside the physical region with the latter close 
to the boundary $u_{1-} \le u_1 \le u_{1+}$.  

\begin{figure}[t]
   \centering
   \vspace{-4.cm}
   \includegraphics[width=1.\textwidth, clip=true]{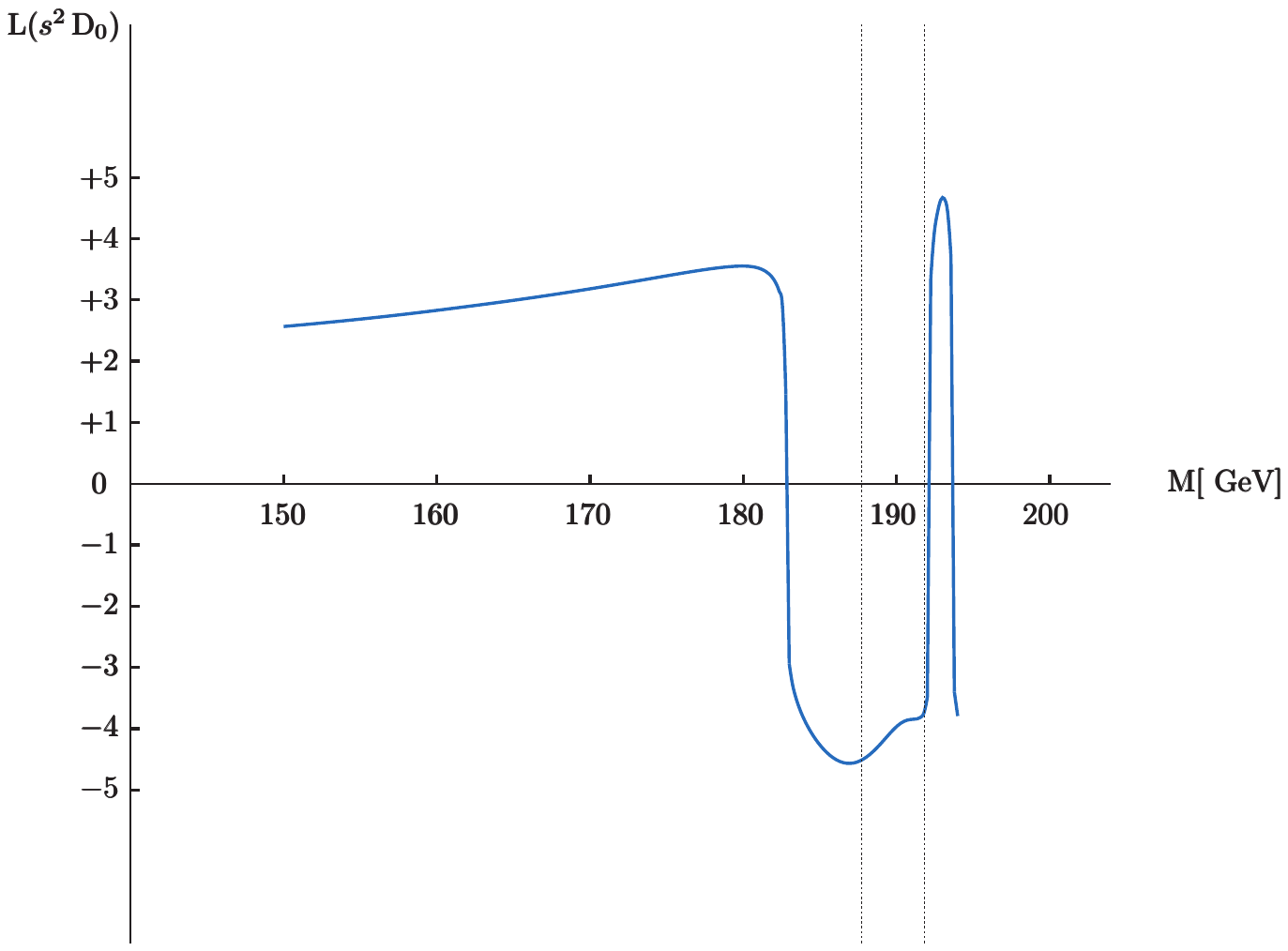}
\vspace{-9.cm}
\caption[]{The real part of $s^2\,\mrD_0$ for $\Pg(Q) \to \PQb(q_1) + \PAQb(q_3) + \PX(q_2)$ at the point of \eqn{refP}.
$\mrL(x)$ is defined in \eqn{logmod}.}
\label{ATfig15}
\end{figure}

\begin{figure}[tb]
   \centering
   \vspace{-4.cm}
   \includegraphics[width=1.\textwidth, clip=true]{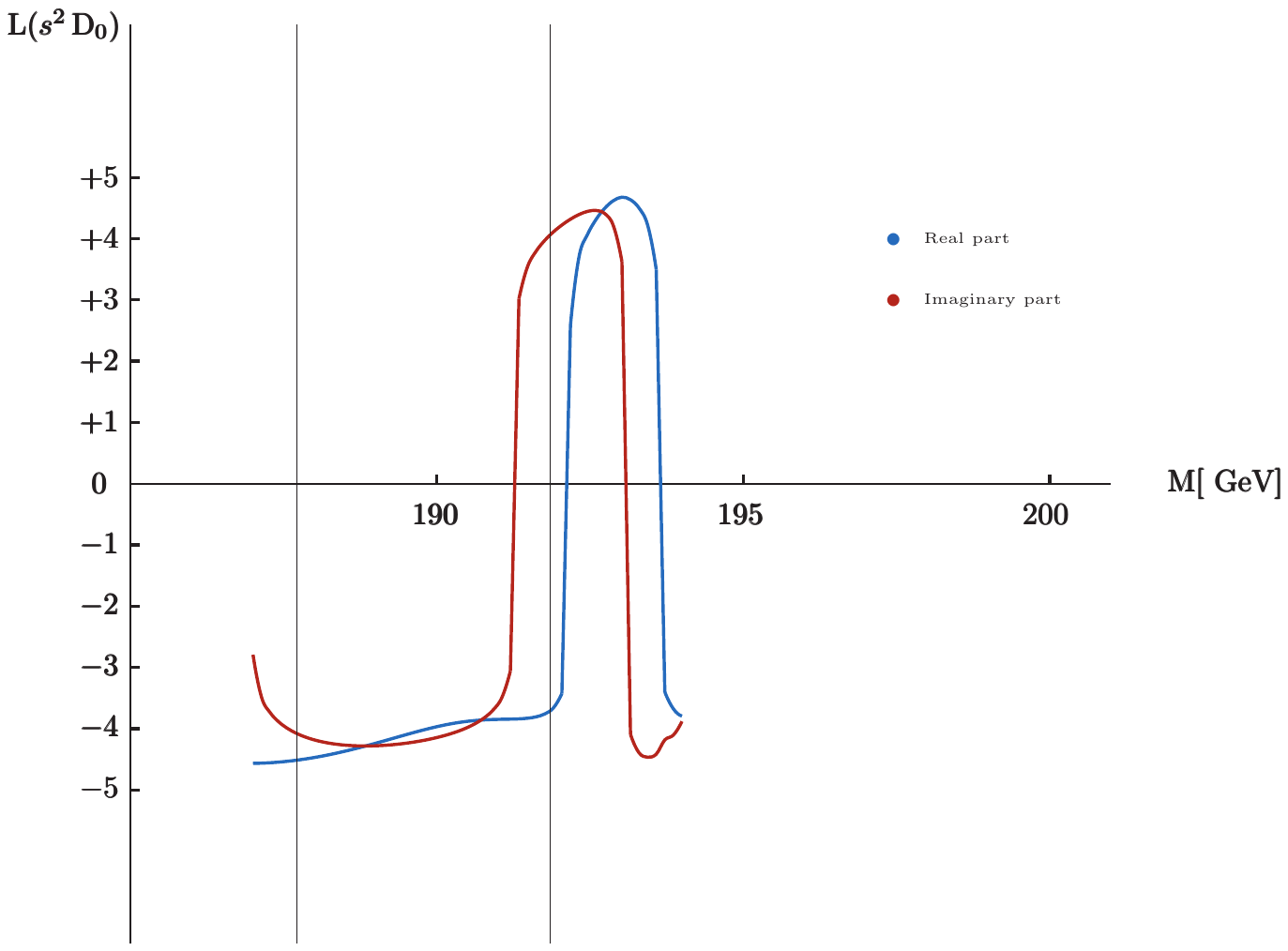}
\vspace{-9.cm}
\caption[]{Blow up of \refig{ATfig15}.}
\label{ATfig16}
\end{figure}

In \refig{ATfig15} we plot $\mrL(s^2\,\Re\,\mrD_0)$ (\eqn{logmod}) corresponding to
$\Pg(Q) \to \PQb(q_1) + \PAQb(q_3) + \PX(q_2)$ at the point of \eqn{refP}. A blow up of the same figure is shown
in \refig{ATfig16}, including the imaginary part.

\begin{figure}[tb]
   \centering
   \vspace{-4.cm}
   \includegraphics[width=1.\textwidth, clip=true]{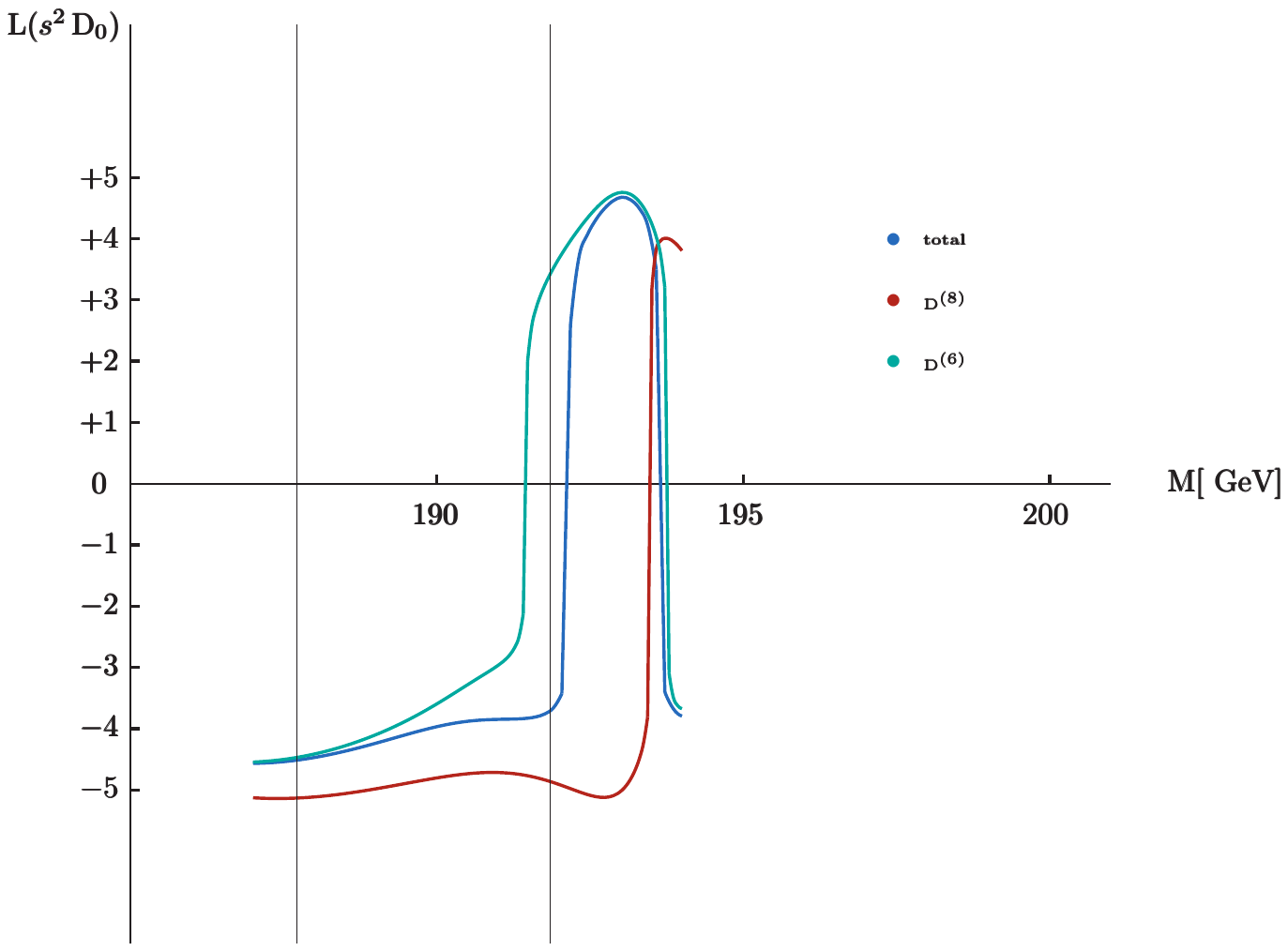}
\vspace{-9.cm}
\caption[]{Different components for $\Re\,\mrD_0$ as explained in \eqn{splitD0}.}
\label{ATfig17}
\end{figure}

To understand the behavior of $\mrD_0$ we split the function as follows: $\mrV(x_1\,,\,x_2\,,\,x_3)$ is the
quadratic form for the box; introduce triangles
\bq
\mrT = \sum_{i=0}^3\,\lpar X_i - X_{i+1} \rpar\,\dsimp{2}\,\mrV^{-1}\lpar \widehat{i\;i+1} \rpar \spc
\eq
where $X_0 = 1$, $X_4= 0$ and $(\widehat{0\;1}) = (1\,,\,x_1\,,\,x_2)$ \etc are contractions.
Therefore we obtain
\bq
\mrD_0(\mrdim=4) = - \frac{1}{2\,\Delta_4}\,\Bigl[ \mrD_0(\mrdim=6) - \mrT \Bigr] =
 - \frac{1}{2\,\Delta_4}\,\Bigl\{ \frac{3}{2\,\Delta_4}\,\Bigl[ \mrD_0(\mrdim=8) - \frac{1}{2}\,\mrT \Bigr]\,
- \mrT \Bigr\} \spc
\eq
and plot
\bq
\mrD_0(\mrdim=4) \spc \quad
\mrD^{(6)}_0(\mrdim=4) = - \frac{1}{2\,\Delta_4}\,\mrD_0(\mrdim=6) \spc \quad
\mrD^{(8)}_0(\mrdim=4) = - \frac{3}{4\,\Delta^2_4}\,\mrD_0(\mrdim=8) \spc
\label{splitD0}
\eq
where the $\mrdim=8$ part is
\bq
\mrD_0(\mrdim=8) = \dsimp{3}\,\Bigl[ \ln V(x_1\,,\,x_2\,,\,x_3) + \frac{2}{3} \Bigr] \spp
\eq
The $3$ components introduced in \eqn{splitD0} are shown in \refig{ATfig17}
for $\Pg(Q) \to \PQb(q_1) + \PAQb(q_3) + \PX(q_2)$ at the point of \eqn{refP}.

\begin{figure}[tb]
   \centering
   \vspace{-4.cm}
   \includegraphics[width=1.\textwidth, clip=true]{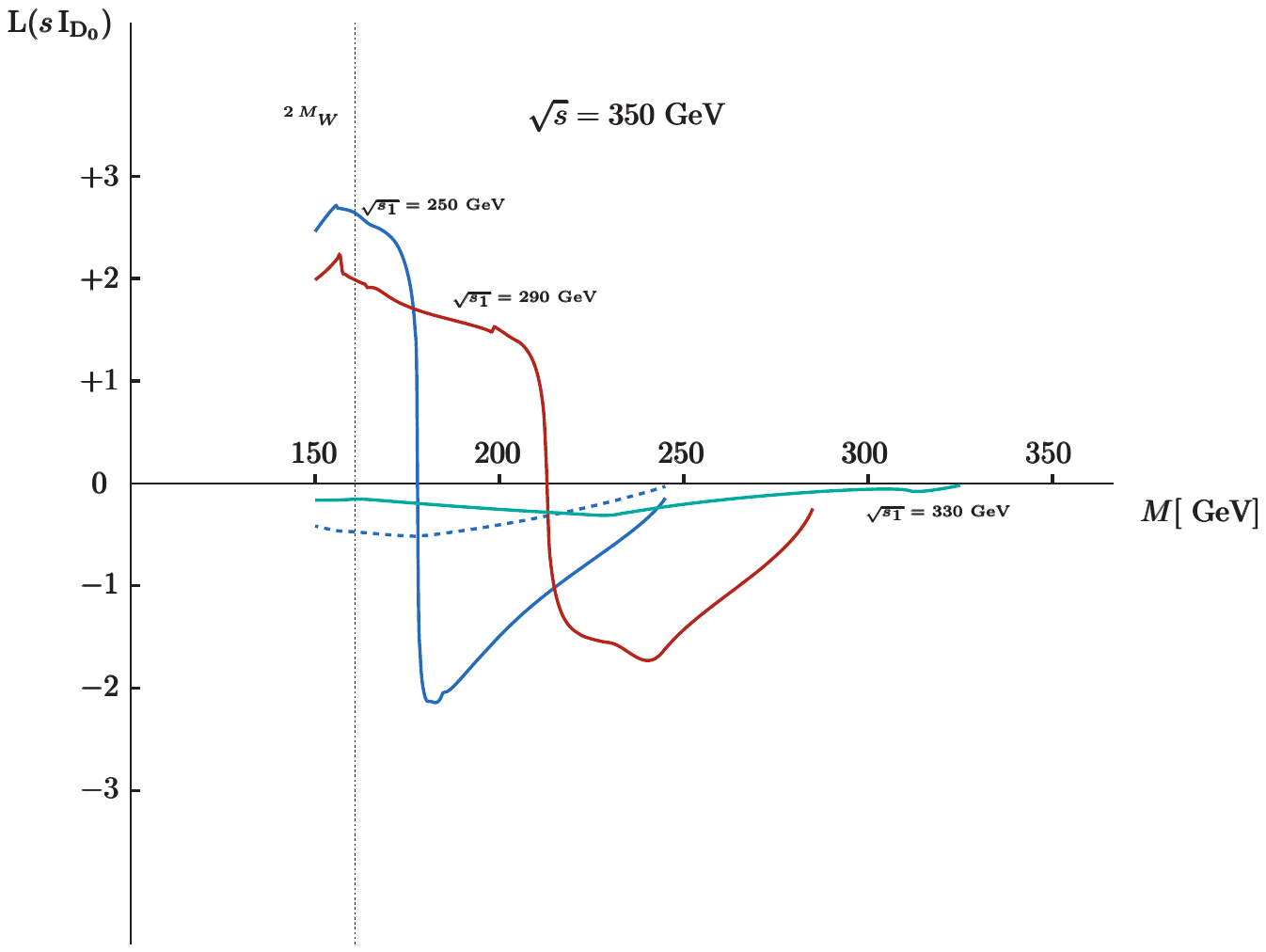}
\vspace{-9.cm}
\caption[]{$\mrD_0\,$-function integrated over $u_1$ at $\sqrt{s} = 350\UGeV$ and for three values of $s_1$.}
\label{ATfig18}
\end{figure}

Finally we introduce
\bq
\mrI_{\mrD_0} = \int_{u_{1-}}^{u_{1+}}\,du_1\,\Re\,\mrD_0 \spc
\eq
where $u_{1\pm}$ are given in \eqn{u1lims}. The corresponding function $\mrL(s\,\mrI_{\mrD_0})$ (\eqn{logmod}) is shown in
\refig{ATfig18} as a function of $\mrM$ for three values of $s_1$ and $\sqrt{s} = 350\UGeV$. Here the 
$\mrD_0\,$-function correspond to $m_1 = m_4 = m_{\PQt}$ and $m_2 = m_3 = \mw$. To show the impact of the AT
we also plot (dashed blue line) the $\mrD_0\,$-function corresponding to $m_1 = m_4 = \mw$ and
$m_2 = m_3 = m_{\PQt}$, a configuration where the AT is absent. 

\paragraph{More boxes} \hspace{0pt} \\
Other examples involving box diagrams are $\Pg + \Pg (\Pep + \Pem) \to \PWp + \PWm + \PH$, with an on-shell Higgs boson
and, at least, one off-shell $\PW$. Here $\Pg + \Pg \to \Pg$ or $\Pep + \Pem \to \PZ/\PGg$ with $\Pg(\PZ/\PGg)$ 
attached to a box where the other external lines are $\PW\PW$ and the Higgs boson (coupled to $\PQb$ internal
lines). 
\subsection{Anomalous threshold and gauge invariance \label{ATGI}}
Given the amplitude for a process supporting an anomalous threshold within the physical region we would like
to split it into two components, \ie $\mrA_{\AT} + \mrA_{\reg}$. In order to have a meaningful result we must
discuss gauge invariance. Consider an of-shell gluon producing a $\PAQb \PQb$ pair and an heavy Higgs, \eg the one 
in the singlet extension of the SM; the full process will be $\Pg + \Pg \to \PAQb + \PQb + \PH$. To discuss gauge 
invariance we work in the $\mrR_{\xi}\,$-gauge where propagators are
\bq
\PW \;\to\; \frac{1}{p^2 + \mws}\,\lpar \delta_{\mu\nu} + \frac{p_{\mu}\,p_{\nu}}{\mws} \rpar -
\frac{p_{\mu}\,p_{\nu}}{\mws\,(p^2 + \xi^2\,\mws)} \spc
\quad
\upphi \;\to\; \frac{1}{p^2 + \xi^2\,\mws} \spc
\eq
where $\upphi$ is a Higgs-Kibble ghost. Consider the four diagrams, family a), shown in \refig{ATfig3}: it is easy 
to show that the corresponding AT can be physical. Next we perform a ``scalarization'' of the amplitude which gives 
a collection of $\mrD_0\,$-functions, $\mrC_0\,$-functions and $\mrB_0\,$-functions. 

\begin{definition}[$\mrD_0\,$-approximation]
It is easy to see that the $D_0$ part of the amplitude is $\xi\,$-independent, \ie $\mrD_0\,$-functions depending 
on $\xi^2\,\mws$ cancel. Therefore we could define $\mrA_{\AT}$ in terms of scalar boxes, including the rest in 
$\mrA_{\reg}$. 
\end{definition}

\begin{figure}[t]
\begin{minipage}{1.\textwidth}
   \includegraphics[width=0.65\textwidth, clip=true]{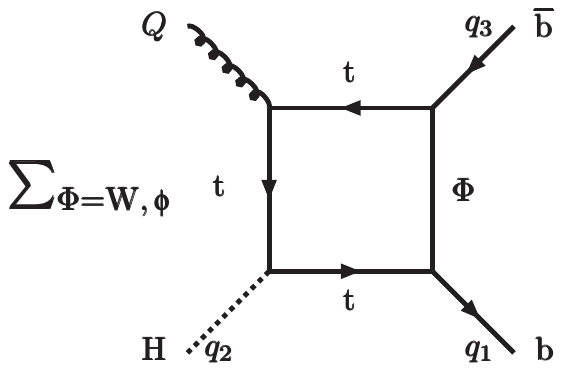}
   \hspace{-6.cm}
   \includegraphics[width=0.65\textwidth, clip=true]{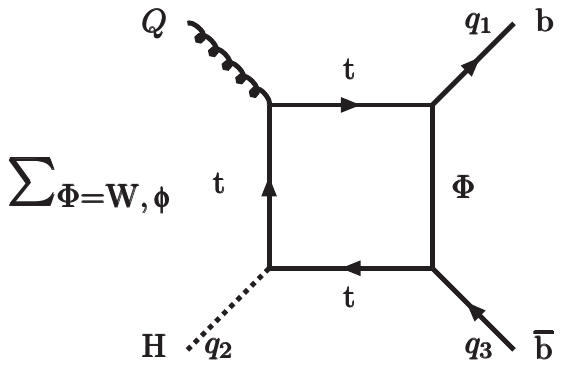}
\vspace{-10.cm}
\caption[]{Family b) for $\Pg(Q) \to \PQb(q_1) + \PAQb(q_3) + \PH(q_2)$.}
\label{ATfig89}
\end{minipage}
\end{figure}

\begin{definition}[Minimal subset]
Alternatively, we can search for $\mrS_{\AT}$, the minimal subset of diagrams which is $\xi\,$-independent, 
satisfies Ward{-}Slavnov{-}Taylor identities (if applicable) and supports a physical-region AT. The corresponding procedure 
can be visualized as follows: scalarization produces triangles which are contractions of the original box, \ie are 
obtained by shrinking one line of the box to a point; therefore, we must add family b), \ie all boxes that give the 
same set of contractions (see \refig{ATfig89}), and family c), \ie all $\PQb{-}\PQb{-}\PH$ (pure) triangles with 
a gluon coupled to $\PQb$ current. 
\end{definition}

\subsubsection{Renormalization}
As has been said many times our goal is not to perform the full calculation, therefore renormalization has to
be understood in the $\MSB$ scheme with the renormalization scale set at the highest scale in the process under
consideration. 
\subsubsection{ \texorpdfstring{$\pi^2$}{} enhancement \label{penh}}
A $\mrC_0\,$-function is given by the integral
\bq
\mrI = \int_0^1 dx_1\,\int_0^{x_1} dx_2\,\Bigl[ \sum_{i,j=1,2}\,(x_1 - X_i)\,\mrH_{ij}\,(x_j - X_ j) + 
\Delta_3 \Bigr]^{-1} \spp
\eq
In general, in the complex mass scheme $\Delta_3, X_{1,2} \in \Cf$.
There are configurations where $\mid \Delta_3 \mid \muchless 1$, although far from the AT, and where
$\mid \mrH_{11} \mid \muchless 1$ (or $\mid \mrH_{22} \mid \muchless 1$) with $0 < \Re X_2 < 1$ (or $0 < \Re X_1 < 1$
but not both). The condition for $\mrH_{ii}$ is always satisfied when we have light external particles. Therefore, 
choosing the case $\mid \mrH_{22} \mid \muchless 1$, we can write
\bqa
\mrI &\approx& \int_0^1 dx_1\,\int_0^{x_1} dx_2 \,\Bigl[ \mrH_{11}\,(x_1 - X_1)^2 +
     2\,\mrH_{12}\,(x_1 - X_1)\,(x_2 - X_2) \Bigr]^{-1}
\nl
{}&= & \frac{1}{2}\,\mrH^{-1}_{12}\,\Bigl\{
\Bigl[ \ln(2\,\mrH_{12}\,(X_1 - X_2) - \ln(- 2\,\mrH_{12}\,X_2) \Bigr]\,\ln(1 - \frac{1}{X_1}) +
\quad \mbox{di-logarithms} \quad \Bigr\} \spp
\eqa
With $\Re X_1 \in [0\,,\,1]$ and $\mid \Im X_1 \mid \muchless 1$ but $\Im X_1 > 0$, we obtain
\bq
\mrI \approx \frac{i\,\pi}{2}\,\mrH^{-1}_{12}\,\Re\,
\Bigl[ \ln(2\,\mrH_{12}\,(X_1 - X_2) - \ln(- 2\,\mrH_{12}\,X_2) \Bigr] + \quad \mbox{rest} \spp
\eq
This $\mrC_0\,$-function will be part of the one-loop corrections to a given LO amplitude. Since $\mrA_{\myLO}$
is in general real this ``$i\,\pi$'' term will not interfere but the NLO amplitude squared will receive a ``$\pi^2$''
enhancement. As we will see in the following sections this is often the case, resulting in NNLO corrections 
(the NLO squared) which are much larger than the NLO ones. Clearly, only a complete calculation can decide the fate 
of these ``$\pi^2$" terms. Note that this enhancement should not be confused with the ``pinch'', the latter
requiring both $X_1$ and $X_2$ to be inside $[0\,,\,1]$.
\subsection{The process  \texorpdfstring{$\Pg + \Pg  \to  \PAQb + \PQb + \PX$}{}}
The full process to be studied will be
\bq
\Pg(p_1) + \Pg(p_2) \to \PQb(q_1) + \PX(q_2) + \PAQb(q_3) \spc
\eq
requiring the following set of invariants:
\bq
s = - P^2 = - (p_1 + p_2)^2 \spc \quad
s_1 = - (P - q_1)^2 \spc \quad
u_1 = - (P - q_2)^2 \spc \quad
t_0 = - (p_1 - q_1)^2 \spc \quad
t_1 = - (p_1 - q_2)^2 \spp
\label{gghbbinv}
\eq
Here $\PX$ is a scalar state of invariant mass $\mrM$. Therefore it can represent some heavy BSM Higgs boson, a 
pair of SM Higgs boson ($\Pg \Pg \to \PAQb \PQb \PH \PH$), a pair of $\PW$ bosons 
($\Pg \Pg \to \PAQb \PQb \PWp \PWm$). 

For real masses we observe that the region of phase space
containing ATs becomes smaller and smaller when $s$ increases and disappears for $\sqrt{s}$ approximately
greater that $438.5\UGeV$. This is not the case for Peierls zeros; we have analyzed those zeros which are
inside the physical phase space and for which 
\bq 
0 \;<\; \Re\,X_3 \;<\; \Re\,X_2 \;<\; \Re\,X_1 \;<\; 1 \spp
\eq
To give an example we consider $\sqrt{s} = 865\UGeV$ and $\sqrt{s_1} = 516.21\UGeV$. There are four solutions to
$\Delta_4 = 0$ and one of them satisfies the requested conditions, corresponding to $\mrM = 308.06\UGeV$ and
$\sqrt{u_1} = 556.94\UGeV$ with
\bq
X_3 = 0.186 + i\,0.977\,10^{-3} \spc \quad
X_2 = 0.250 - i\,0.171\,10^{-2} \spc \quad
X_1 = 0.625 + i\,0.793\,10^{-3} \spp
\eq
The evaluation of loop integrals which are regular when the corresponding BST factor is zero requires an
additional comment: when internal masses are real the solution has been described in \Bref{Ferroglia:2002mz}. 
For instance, consider the three-point function, characterized by a polynomial
\bq
\mrV(x_1\,,\,x_2) - i\,0 = \Bigl[ \mrV_0(x_1\,,\,x_2) - i\,0 \Bigr] + \lpar \Delta_3 - i\,0 \rpar \spp
\eq
In section~4.4.1. of \Bref{Ferroglia:2002mz} it is found that one can write
\bq
\mrC_0 = \frac{1}{\Delta_3}\,\Bigl[
\dsimp{2}\,\ln\frac{\mrV}{\mrV_0} - \frac{1}{2}\,\sum_{i=0}^{2}\,(X_i - X_{i+1})\,
\dsimp{1}\,\ln\frac{\mrV(\widehat{i\;i+1})}{\mrV_0(\widehat{i\;i+1})} \Bigr] \spp
\label{C0sub}
\eq
When complex masses are introduced the imaginary part of $\mrV$ has always the same sign, but this is not
the case for $\mrV_0$; furthermore, $\Im \mrV$ and $\Im \mrV_0$ have, in general, different signs so that
one cannot reconstruct terms like $\ln(\mrV/\mrV_0)$ as done in \eqn{C0sub}. Therefore, we will write
\bq
\mrV(x_1\,,\,x_2\,;\,\{\mrI\}) = \mrV_0(x_1\,,\,x_2\,;\,\{\mrI\}) + \Delta_3(\{\mrI\}) \spc
\eq
where $\{\mrI\}$ is the set of invariants describing the process. When
\bq
\Delta_3(\{\overline{\mrI}\}) = 0 \spc
\eq
we will subtract, with a double BST algorithm, $\mrV_0(x_1\,,\,x_2\,;\,\{\overline{\mrI}\})$ taking care of
reconstructing $\ln(\mrV/\mrV_0)$ only when $\mid \Delta_3 \mid$ is small enough and the imaginary parts of
numerator and denominator have the same sign. Summarizing: let $\overline{X}_i = X_i(\{\overline{\mrI}\})$,
we will use
\bqa
\Bigl[ 1 + \frac{1}{2\,(\mu + 1)}\,(x - X)^{\mathrm{t}}\,\partial_x \Bigr]\,\mrV^{\mu+1}(x_1\,,\,x_2\,;\,\{\mrI\}) &=&
\Delta_3(\{\mrI\})\,\mrV^{\mu}(x_1\,,\,x_2\,;\,\{\mrI\}) \spc  
\nl
\Bigl[ 1 + \frac{1}{2\,(\mu + 1)}\,(x - {\overline{X}})^{\mathrm{t}}\,\partial_x \Bigr]\,
\mrV^{\mu+1}_0(x_1\,,\,x_2\,;\,\{{\overline{\mrI}}\}) &=& 0 \spc
\eqa
subtract the two equations and integrate by parts.

For $\Pg + \Pg \to \PAQb + \PQb + PX$ where $\PX$ is an off-shell SM Higgs boson or an on-shell BSM Higgs boson the  
first three invariants in \eqn{gghbbinv} are
\bq 
s_{1-} \le s_1 \le s_{1+} \spc \dots \spc t_{0-} \le t_0 \le t_{0+} \spc
\eq
and we introduce new, scaled variables,
\bq
s_1 = (s_{1+} - s_{1-})\,y_1 + s_{1-} \spc \dots \spc t_0 = (t_{0+} - t_{0-})\,y_3 + t_{0-} \spp
\eq
For the last invariant in \eqn{gghbbinv} we use $\lambda_1 = \lambda(s\,,\,s_1\,,\,m^2_{\PQb})$ and
$\lambda_3 = \lambda(s\,,\,u_1\,,\,\mrM^2)$ and define
\bq
\xi = (s - s_1 + 2\,t_0 - m^2_{\PQb})\,\lambda^{-1/2}_1 \spc
\qquad
\eta = \Bigl[ 2\,s\,(s_1 + \mrM^2) - (s + s_1)\,(s - u_1 + \mrM^2) - m^2_{\PQb}\,(s + u_1 - \mrM^2)\Bigr]\,
(\lambda_1\,\lambda_3 )^{-1/2} \spp
\eq
Next we introduce
\bq
\mrz = \pi\,(y_4 - \frac{1}{2}) \spc \qquad y_4 \in [0\,,\,1] \spc
\eq
and obtain
\bq
t_1 = \frac{1}{2}\,\Bigl\{\Bigl[ (\xi\,\eta)^{1/2}\,\sin(\mrz) + 
\xi\,\eta \Bigr]\,\lambda^{1/2}_3 - s + u_1 + \mrM^2\Bigr\} \spp
\eq
The object of interest (fully differential) is
\bq
\frac{d^4\,\sigma}{dy_1\,\dots\,dy_4} = \frac{1}{32\,\pi^3}\,\frac{s_{1+} - s_{1-}}{512\,s_1\,s^2}\,
( \lambda_1\,\lambda_2 )^{1/2}\,\sum_{s,c}\,\mid \mrA \mid^2 \spc
\eq
where $\lambda_1 = \lambda(s\,,\,s_1\,,\,m^2_{\PQb})$ and $\lambda_2 = \lambda(s_1\,,\,\mrM^2\,,\,m^2_{\PQb})$.
The sum is over spin and color, $\mrA$ is the amplitude. Finally, 
\bq
s_{1-} = (\mrM + m_{\PQb})^2 \spc \qquad s_{1+} = (\sqrt{s} - m_{\PQb})^2 \spp
\eq
Both LO and interference with one-loop corrections are proportional to $m^2_{\PQb}$ while one-loop squared
survives the limit $m_{\PQb} \to 0$. Loop effects are suppressed by a factor $g^2/(16\,\pi^2) = 0.0027$ which, however
is of the same order of magnitude of the Yukawa suppression $m^2_{\PQb}/\mws \approx 0.0021$. Therefore, we
expect one-loop squared to be of the same order of the interference.

We selected $\sqrt{s} = 865\UGeV$ and $\mrM = 308.06\UGeV$. In the four-dimensional $y\,$-space there are 
trajectories feeling the presence of the Peierls zero, for instance, with $y_3 = y_4 = 0.3$ and $y_2 = 0.6$ 
we have (NNLO is up to one-loop squared)
\[
\begin{array}{lccc}
{} \;\;&\;\; y_1 = 0.25 \;\;&\;\; y_1= 0.2626 \;\;&\;\; y_1 = 0.27 \\
\delta^{\AT}_{\myNLO}[\%] \;\;&\;\; 8.14 \;\;&\;\; 3.66 \;\;&\;\; 8.41 \\
\delta^{\AT}_{\myNNLO}[\%] \;\;&\;\; 87.03 \;\;&\;\; 370.25 \;\;&\;\; 84.32
\end{array}
\]
It is worth noting the very large NNLO effect, induced by the ``$\pi^2$' terms which originates from triangles, as
explained in \autoref{penh}. 
For comparison we give $\delta^{\AT}_{\myNNLO}$ for different values of $\sqrt{s}$ with $\mrM$ fixed and 
$y_1= 0.2626$, $y_2 = 0.6$, $y_3 = y_4 = 0.3$:
\[
\begin{array}{lccc}
\sqrt{s}[\UGeV] \;\;&\;\; 350 \;\;&\;\; 550 \;\;&\;\; 865 \\
\delta^{\AT}_{\myNNLO} \;\;&\;\;  - 0.68 \;\;&\;\; 10.56 \;\;&\;\; 370.25  
\end{array}
\]
showing the combined effect of the Peierls zero and the ``$\pi^2$" terms.
Unfortunately, when more ($y$) integrations are performed the effect becomes less and less visible; for instance in the
two-dimensional distribution, $y_1{-}y_2$, trajectories are almost flat in the $y_1(y_2)$ direction.
This seems to be a general result: for processes requiring more and more invariants the fully inclusive observables become
less and less sensitive to ATs.
\subsection{The process  \texorpdfstring{$\Pg + \Pg \to \PAQb + \PQb + \PH + \PH$}{}}
\begin{figure}[tb]
   \centering
   \vspace{-4.cm}
   \includegraphics[width=1.\textwidth, clip=true]{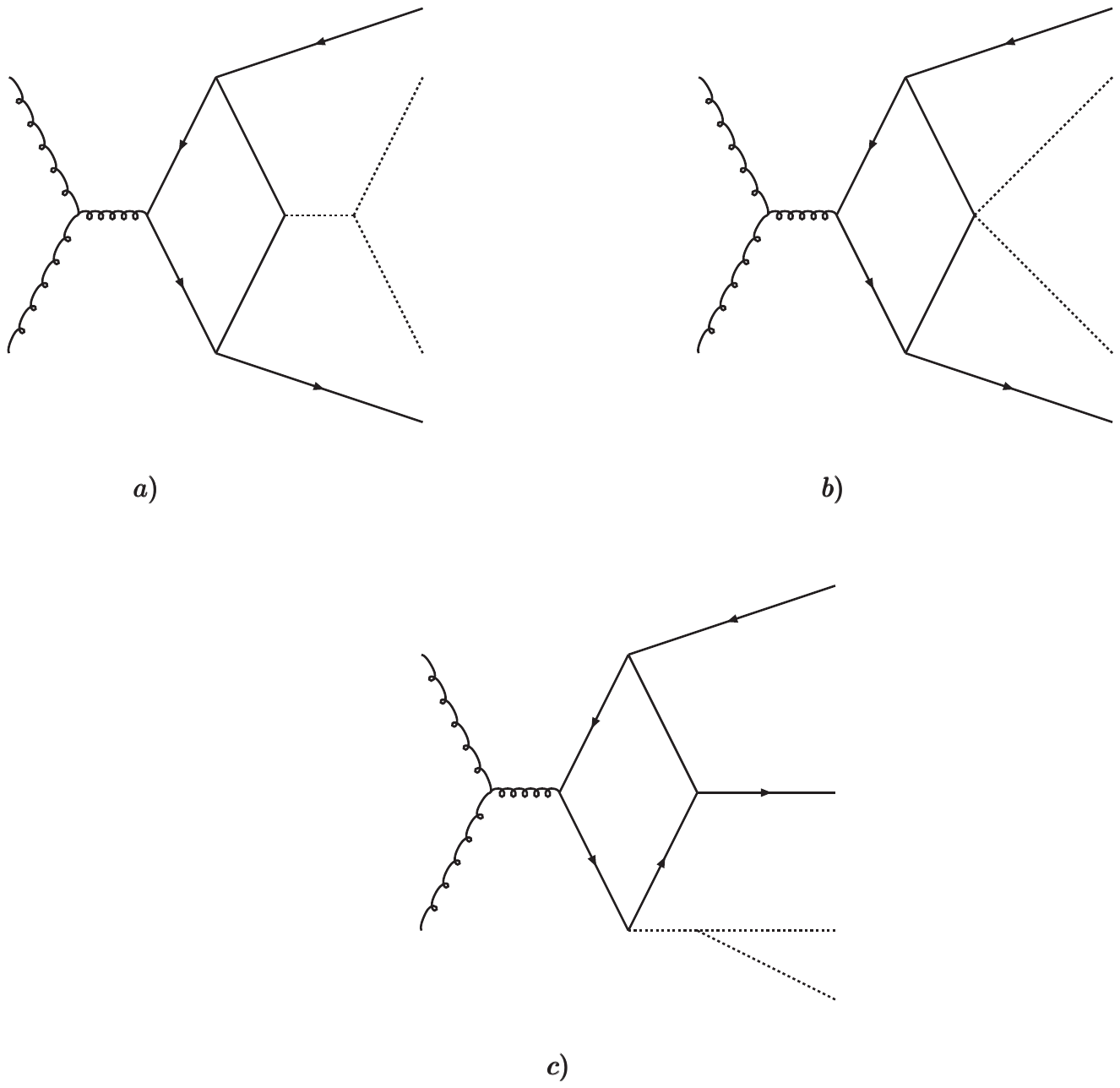}
\vspace{-11.cm}
\caption[]{Diagrams contributing to $\Pg + \Pg \to \PAQb + \PQb + \PH + \PH$.}
\label{ATfig19}
\end{figure}

\begin{figure}[tb]
   \centering
   \vspace{-4.cm}
   \includegraphics[width=1.\textwidth, clip=true]{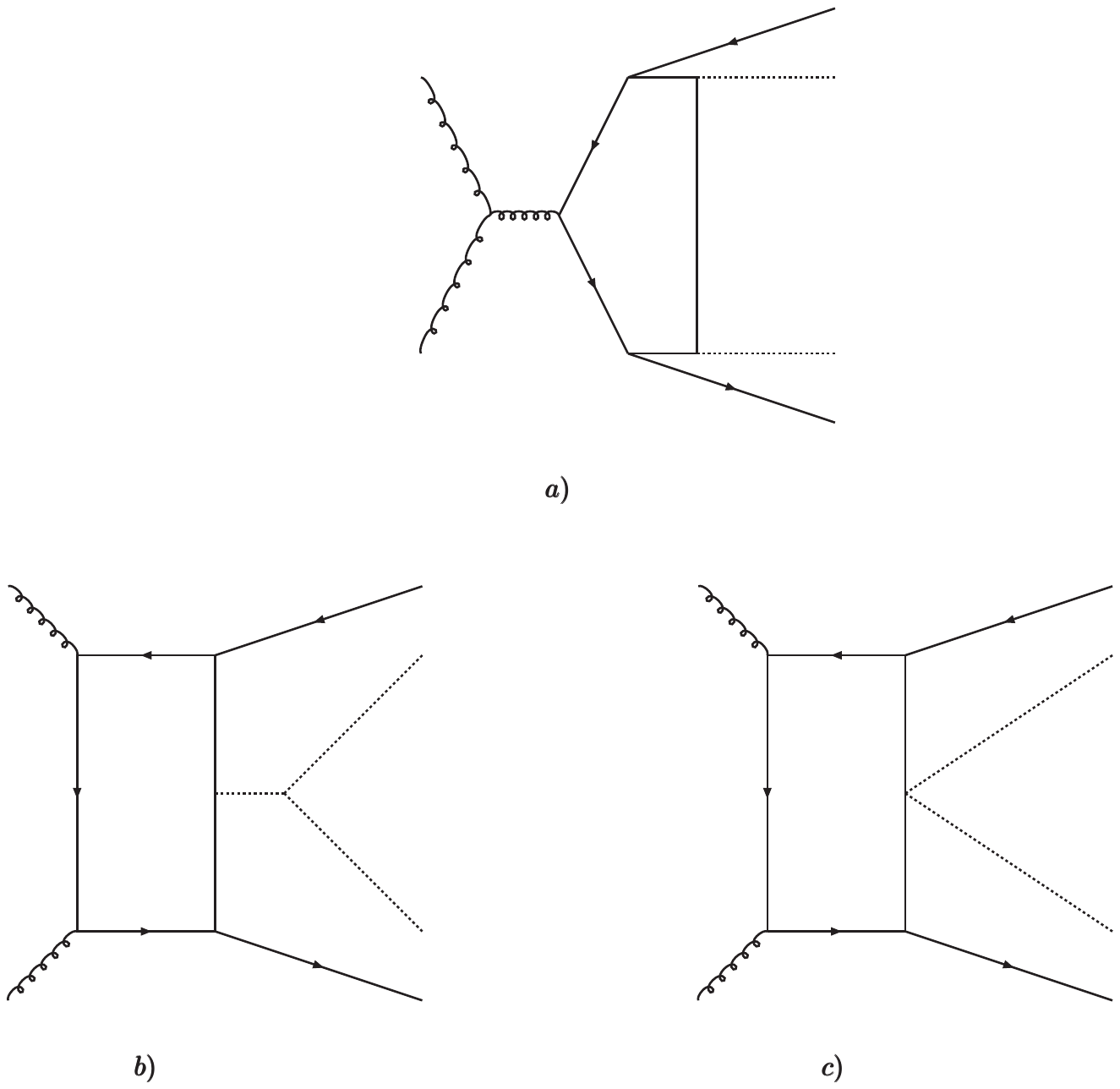}
\vspace{-11.cm}
\caption[]{Diagrams contributing to $\Pg + \Pg \to \PAQb + \PQb + \PH + \PH$.}
\label{ATfig20}
\end{figure}

There are several classes of diagrams contributing to the process.
In \refig{ATfig19} diagrams of class a) have an AT inside the physical region; diagrams of class b) do not have an AT 
but are needed for gauge invariance when we move beyond the $\mrD_0\,$-approximation. Diagrams in \refig{ATfig20}
include pentagons which do not have a leading singularity in the physical region but show a subleading one
(box driven) obtained by shrinking one line of the pentagon to a point.

\section{Pentagons \label{penta}}
As shown in \Bref{Melrose:1965kb} there is no discontinuity associated with the LLS
of the pentagon. This is why \Bref{Cutkosky:1960sp} refers to the singularity as a pole, \ie
$\mrE_0 \sim 1/\Delta_5$, as shown in \Bref{Ferroglia:2002mz}.

\begin{figure}[tb]
   \centering
   \vspace{-2.cm}
   \includegraphics[width=1.\textwidth, clip=true]{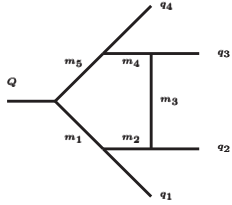}
\vspace{-15.cm}
\caption[]{Pentagon diagram: the general case with arbitrary internal and external masses.}
\label{ATfig13}
\end{figure}

Consider the diagram of \refig{ATfig13} with $Q^2 = - s$, $q^2_i= - \mrM^2_i$. 
The corresponding process is described in terms of the following set of invariant:
\bq
s_1 = - (Q - q_1)^2 \spc \quad
s_2 = - (Q - q_1 - q_2)^2 \spc \quad
u_1 = - (Q - q_2)^2 \spc \quad
u_2 = - (Q - q_3)^2 \spc \quad
t_2 = - (Q - q_2 - q_3)^2 \spp
\eq
Their limits, the physical region, are:
\bq
(\mrM_2 + \mrM_3 + \mrM_4)^2 \le s_1 \le (\sqrt{s} - \mrM_1)^2 \spc \qquad
(\mrM_3 + \mrM_4)^2 \le s_2 \le (\sqrt{s_1} - \mrM_2)^2 \spc
\eq
and $u_{1-} \le u_1 \le u_{1+}$ \etc where the limits can be written as $u_{1\pm} = u_{10} \pm du_{1}$ \etc
with the following explicit expressions (see \Bref{Kumar:1970cr} for details),
\bqa
{}&{}& 
u_{10} = s + \mrM^2_2 - \frac{1}{2\,s_1}\,(s_1 - s_2 + \mrM^2_2)\,(s + s_1 - \mrM^2_1) \spc \quad
\Delta u_1 = \frac{1}{2\,s_1}\,\lambda_1\,\lambda_2 \spc
\nl
{}&{}&
u_{20} = s + \mrM^2_3 - \frac{1}{2\,s_2}\,(s_2 + \mrM^2_3 - \mrM^2_4)\,(s + s_2 - s^\prime_2) \spc \quad
\Delta u_2 = \frac{1}{2\,s_2}\,\lambda_3\,\lambda_4 \spc
\nl
{}&{}&
t_{20} = u_1 + \mrM^2_3 - \frac{1}{2\,s}\,(s - u_2 + \mrM^2_3)\,(s + u_1 - \mrM^2_2) - 
         \frac{1}{2}\,\frac{\xi_2 \eta_2}{s}\,\lambda_5\,\lambda_6 \spc \quad
dt_2 = \frac{1}{2\,s}\,(1 - \xi^2_2)^{1/2}\,(1 - \eta^2_2)^{1/2}\,\lambda_5\,\lambda_6 \spp
\eqa
Variables $\lambda_1,\,\dots\,,\lambda_6$ are defined in terms of the following K\"allen lambda functions:
\bqa
\lambda_1 &=& \lambda^{1/2}(s\,,\,s_1\,,\,\mrM^2_1) \spc \quad
\lambda_2 = \lambda^{1/2}(s_1\,,\,s_2\,,\,\mrM^2_2) \spc \quad
\lambda_3 = \lambda^{1/2}(s\,,\,s_2\,,\,s^\prime_2) \spc 
\nl
\lambda_4 &=& \lambda^{1}{2}(s_2\,,\,\mrM^2_3\,,\,\mrM^2_4) \spc \quad
\lambda_5 = \lambda^{1/2}(s\,,\,u_1\,,\,\mrM^2_2) \spc \quad
\lambda_6 = \lambda^{1/2}(s\,,\,u_2\,,\,\mrM^2_3) \spp
\eqa
We have also introduced additional variables 
\bq
s^\prime_2 = s - s_1 + s_2 - u_1 + \mrM^2_1 + \mrM^2_2 \spc \qquad t^\prime_1 = \mrM^2_2 \spp
\eq
The variables $\xi_2$ and $\eta_2$ can be found in \Bref{Kumar:1970cr} with full details on the calculation of the
phase space. 

The example to be discussed is as follows:
\bqa
{}&{}& \mrM_1 = m_{\PQb} \spc \quad
\mrM_3 = 0 \spc \quad
\mrM_2 = \mrM \spc \quad
\mrM_4 = m_{\PQb} \spc
\nl
{}&{}& m_1 = m_{\PQt} \spc \quad
m_2 = \mw \spc \quad
m_3 = 0 \spc \quad
m_4 =\mw \spc \quad
m_t = m_{\PQt} \spc
\eqa
where $q_3$ is an electron and we have neglected the mass while line $3$ is a neutrino. In the limit of zero widths
the equation $\Delta_5 = 0$ is quadratic in $t_2$; here $q_2$ is the momentum of a pair, electron{-}photon, so that 
what we are considering
is $\Pg \Pg \to \Pg \to \PAQb \PQb \Pem \Pep \PGg$. An example of physical-region ATs is given in the following table.
\[
\begin{array}{rrrrrr}
\mbox{fixed} &&&&& \mbox{solution} \\
\hline
\mrM_2 [\UGeV] & \sqrt{s_1} [\UGeV] & \sqrt{s_2} [\UGeV]& \sqrt{u_1} [\UGeV]& \sqrt{u_2} [\UGeV]& \sqrt{t_2} [\UGeV]\\
\hline
     97.40   & 260.47   & 139.17   & 201.03   & 275.84  & 122.53 \\
       "     &    "     &    "     & 203.55   & 260.91  & 89.06  \\
       "     &    "     &    "     &    "     & 265.74  & 103.56 \\
       "     &    "     &    "     &    "     & 270.56  & 117.37 \\
       "     &    "     &    "     &    "     & 275.39  & 130.91 \\
       "     &    "     & 147.14   & 213.31   & 266.39  & 121.68 \\
       "     &    "     &    "     &    "     & 271.07  & 138.30 \\ 
       "     &    "     &    "     & 215.30   & 255.98  & 98.96  \\
       "     &    "     &    "     &    "     & 260.89  & 113.19 \\
\hline
\end{array}
\]

The same line of arguments applies to $\Pg \Pg \to \PAQb \PQb \PAQu \PQu \Pg$ or
$\Pg \Pg \to \PAQb \PQb \PAQd \PQd \Pg$, with $\PQu \not= \PQt$.

An interesting question is the following: consider a particle of invariant mass $\sqrt{s}$ ``decaying'' into a 
four-body final state; $\Delta_5$ is a quartic polynomial in $s$ and we look for complex solutions in the
variable $s$ to the equation $\Delta_5 = 0$ with $0 < \Re X_4 < \,\dots\, \Re X_1 < 1$. Since the AT for a
pentagon is a pole this is exactly the situation where the AT could be misinterpreted as the peak due to an
unstable particle. 

Deciding whether it is a resonance would require establishing that the ``pole'' is on the
second (unphysical) sheet. In this case one of the external `masses'', \ie $s$, is complex and  
$\ln^-(\mrz) \not= \ln(\mrz)$ etc. This fact requires a more complicated structure of the analytic continuation of the
original integral including contour deformation, see section~6.6 of \Bref{Passarino:2010qk}, which is beyond the scope 
of this work. Furthermore, the real analytic approach has the merit of working with quantities having a direct 
physical meaning and direct physical intuition is certainly of great help.
The vertex function with three complex external masses has been discussed in \Bref{Bonnevay1961}.

\section{Hexagons \label{hexa}}
For six point functions the leading Landau singularity is a sum of products of the Landau singularities for
the reduced pentagon diagrams~\cite{Melrose:1965kb}.

For $n \ge 6$ all the singularities of the one-loop $n\,$-point functions coincide with the 
singularities of the reduced, down to and including the pentagon, diagrams obtained 
from the main diagram~\cite{Brown:1961aaa,Asribekov:1962tgp}.
This result follows from the well-known property of vanishing of Gram determinants 
if any of its principal minors vanish. A plausible conclusion: at one loop a simple pole is the strongest possible 
singularity.

\section{Two loop diagrams \label{tloops}}
\begin{figure}[t]
   \centering
   \includegraphics[width=1.\textwidth, clip=true]{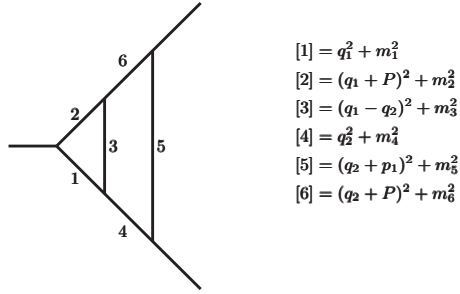}
\vspace{-15.cm}
\caption[]{Two loop diagram. With $[i]$ we denote the inverse propagator for line $i$.}
\label{ATfig14}
\end{figure}

So far we have discussed the leading Landau singularity for one-loop diagrams. In view of the impact of QCD
corrections we would like to understand the behavior of two loop diagrams; this is, by far, more complicated than the 
one loop analysis, especially because we have no simple expression for the BST factor. Therefore, we have to resort
to the set of Landau equations, written in momentum space. The full analysis for two loop triangles has been
given in \Bref{Ferroglia:2003yj} and we will use the relevant results. For a diagram with $l$ loops and $n$
propagators we have $4\,l + n + 1$ conditions on $4\,l + n$ variables; this means that a solution may exist only 
for specific values of external momenta. 
Consider the vertex of \refig{ATfig14}, the Landau equation for this topology are as follows:
\[
\begin{array}{lll}
\alpha_1\,(q^2_1+m^2_1) = 0 \spc \qquad & \qquad
\alpha_2\,((q_1+P)^2+m^2_2) = 0 \spc \qquad & \qquad
\alpha_3\,((q_1-q_2)^2+m^2_3) = 0 \spc    \nl
\alpha_4\,(q_2^2+m^2_4) = 0 \spc \qquad & \qquad
\alpha_5\,((q_2+p_1)^2+m^2_5) = 0 \spc \qquad & \qquad
\alpha_6\,((q_2+P)^2+m^2_6) = 0 \spc
\end{array}
\]
and also
\bqa
\alpha_1 q_{1\mu} + \alpha_2 (q_1+P)_{\mu}
     + \alpha_3 (q_1-q_2)_{\mu} &=& 0 \spc
\nl
- \alpha_3 (q_1-q_2)_{\mu} + \alpha_4\,q_{2\mu}
       + \alpha_5\,(q_2+p_1)_{\mu}  + \alpha_6\,(q_2+P)_{\mu} &=& 0 \spc
\label{landtl}
\eqa

The leading Landau singularity occurs for $\alpha_i \ne 0, \forall i$.
We multiply the two equations \eqn{landtl} by $q_{1\mu}$, $q_{2\mu}$,
$p_{1\mu}$ and $P_{\mu}$, respectively. This gives an homogeneous system of 
eight equations. If all $\alpha_i$ are different from zero we may use
\[
\ba{lll}
q^2_1 = - m^2_1 & \qquad q^2_2 = - m^2_4 \spc & \qquad
\spro{q_1}{q_2} = \frac{1}{2}\,(m^2_3 - m^2_1 - m^2_4) \spc \\
\spro{q_1}{P} = -\,\frac{1}{2}\,( P^2 - m^2_1 + m^2_2) \spc & \qquad
\spro{q_2}{p_1} = -\,\frac{1}{2}\,( p_1^2 - m^2_4 + m^2_5) \spc & \qquad
\spro{q_2}{P} = -\,\frac{1}{2}\,( P^2 - m^2_4 + m^2_6) \spp
\ea
\]
Compatibility requires a set of relations among $P^2, p^2_{1,2}$ and internal masses. If we select $m_3 = 0$ (gluon, photon),
$m_5 = m_{\PQb}$ and the remaining masses equal to $m_{\PQt}$ then a non trivial solution ($\alpha_i \ne 0, \forall i$)
occurs iff
\bq
s = - P^2 = 4\,m^2_{\PQt} \spc \quad
\mrM^2_1 = - p^2_1 = (m_{\PQt} \pm m_{\PQb})^2 \spc \quad
\mrM^2_2 = - p^2_2 = (m_{\PQt} \mp m_{\PQb})^2 \spc 
\eq
\ie exactly at the boundary of phase space. The solution is $\alpha_{1\,,\,3\,,\,4\,,\,5}$ arbitrary and
$\alpha_2 = \alpha_1$, $m_{\PQt}\,\alpha_6 = m_{\PQt}\,\alpha_4 - m_{\PQb}\,\alpha_5$.
This solution includes the case $\alpha_3 = 0$.

It is worth noting that for physical configuration, \ie the real external momenta, the Landau singularities are
on the first (physical) sheet when $\alpha_1 \in [0\,,\,1]$ and
may or may not be on the first (physical) sheet when $\alpha_i \not\in [0\,,\,1]$.
 
If $m_3 = \mz$ we obtain $s = 4\,m^2_{\PQt} - \mzs$ and 
\bqa
{}&{}& 2\,m^2_{\PQt}\,(\mrM^2_1 + \mrM^2_2)^2
- \Bigl[ 8\,(m^2_{\PQt} + m^2_{\PQb})\,m^2_{\PQt} - (3\,m^2_{\PQt} + m^2_{\PQb}) \Bigr]\,(\mrM^2_1 + \mrM^2_2) 
- \mzs\,mrM^2_1\,\mrM^2_2
\nl
{}&-& \Bigl[ (m^2_{\PQt} + m^2_{\PQb})^2 + 4\,m^2_{\PQt}\,(m^2_{\PQt} + 2\,m^2_{\PQb})\,\mzs
+ (m^2_{\PQt} + m^2_{\PQb})\,\mzq
+ 8\,(m^2_{\PQt} + m^2_{\PQb})^2\,m^2_{\PQt} = 0 \spp
\eqa
It is easily seen that the last equation does not have real solutions for $\mrM_{1,2}$ for physical values of
$\mz, m_{\PQt}$ and $m_{\PQb}$.

\section{Special cases \label{Scas}}
As we have seen any one-loop diagram with $\mrN$ external legs is characterized by its BST factor $\Delta_{\ssN}$
defined in \eqn{Gdef}, $\mrN - 1$ coefficients, $X_1\,,\dots\,,X_{\ssN-1}$ defined in \eqn{Xdef} and by a set of 
external masses and Mandelstam invariants.
In the complex mass scheme we have $\Delta_{\ssN}, X_i \in \Cf$. With no loss of generality, we fix $\mrN = 4$ and
consider a general process $Q \to q_1 + q_2 + q_3$ where $Q^2 = - s$ and $q^2_i = - \mrM^2_i$. Let $s_1$ and
$u_1$ be the two independent invariants, defined in \eqn{boxmi}. Consider the following system of equations:
\[
\begin{array}{ll}
\Re \Delta_4 = 0 \quad & \quad \Im \Delta_4 = 0 \\
& \\
\Im X_i = 0      \quad & \quad i=1\,,\dots\,,3
\end{array}
\]
Therefore we have $5$ equations in $6$ unknowns, $s, s_1, u_1, \mrM^2_{1,2,3}$; a solution will give
a surface parametrized by $s_1 = s_1(s), u_1 = u_1(s)$ \etc
If the real part of the $X_i$, evaluated at the solution, is ordered then we may have a pinch singularity even in
the complex mass scheme. The square of the box will be non-integrable.

For the pentagon we have $5$ external masses and $5$ invariants with $4$ $X$ variables, therefore $6$ equations for
$10$ unknowns which, once again, gives a surface of potential singularities. If the $\Re X_i$ are ordered the pentagon
itself may develop a non-integrable singularity, even in the complex mass scheme.

Having or not a singularity depends on which trajectory we follow in phase space, \ie on the order of the two limits,
$\Delta_{\ssN} \to 0$ and $\Im X_i \to 0$.

The fate of these configurations can only be decided on a case-by-case basis; if they appear inside the pysical region
their study must be completed including beam energy spread, parton distribution functions and modelling lossy processes 
(\eg by including a Crystal Ball function). An illustrative example is given in \autoref{iexa}. 

We briefly mention one example, a box corresponding to $Q \to q_1 + q_2 + q_3$ where $Q^2 = - s$ and $q^2_i = - \mrM^2_i$.
There are four internal lines with masses $m_i$; when $m_1 = m_4$ and $m_2 = m_3$ we derive the following result:
given $s_1 = - (q_2 + q_3)^2$ and $u_1 = - (q_1 + q_3)^2$, when
\bq
\mrM_2 = \sqrt{s} - \sqrt{u_1} \spc
\qquad
\mrM^2_3 = - \frac{ s_1\,(u_1 - s) + \mrM^2_2\,(s + s_1 + u_1 - 2\,\mrM^2_1) - \mrM^4_2}{s + \mrM^2_2 - u_1} \spc
\label{scashys}
\eq
it follows that $\forall i\,,\,\Im\,X_i = 0$ and $\Im\mrC_4 = 0$. However, on the hypersurface defined by
\eqn{scashys} we have $\mrG_4 = 0$. Therefore, the box is a linear combination of four triangles divided by
$\mrC_4$. 

We can now solve for $\Re\,\mrC_4 = 0$ where $\Re\,\mrC_4$ is a real polynomial of fourth degree in 
$\mrM^2_1$; the solutions, $\mrM^2_1(s_1\,,\,u_1)$, are the surfaces where $\mrC_ 4 \to 0$ after $\mrG_4 \to 0$
(this case in discussed in \ref{SB}).
We are looking for solutions where $\mrM^2_{1,2,3} \in \Rf_{>0}$ and where $s_1$ and $u_1$ are within their boundaries
for fixed $s$. These conditions are very difficult to satisfy and only in few cases we have found real positive
(squared) masses and $\mrM_2 + \mrM_3 \le \sqrt{s_1} \le \sqrt{s} - \mrM_1$. As far as $u_1$ is concerned we have found
(with a scan in $s{-}s_1{-}u_1$) that $u_1$ is always larger than its (physical) upper bound, even if in a very limited
number of cases the difference $\sqrt{u_1} - \sqrt{u_{1+}}$ can be as small as $1\,\UeV$.
\subsection{Folding  the AT \label{iexa}}
Consider a process $\Pep \Pem \to \PX$; in the so-called ``radiator approach'' the hard scattering cross section is
convoluted with initial state QED radiation,
\bq
\sigma(s) = \int_0^1 d\mrz\,\mrH(\mrz,s)\,{\widehat{\sigma}}((1 - \mrz)\,s) \spc
\eq
where $s = 4\,E^2$, $E$ being the beam energy. 
We are not concerned here with the exact form of the radiator, it will be enough to assume the so-called virtual-soft
approximation where
\bq
\mrH = \mrH_0\,\beta\,\mrz^{\beta - 1} \spc \qquad 0 < \beta \muchless 1 \spp
\eq
The simple example we have in mind is
\bq
\mrF(s) = \int_0^1 d\mrz\,\mrH(\mrz\,,\,s)\,f(s) \spc
\qquad
f(s)= \int_0^1 dx\,\chi^{-3/2}(x\,,\,s) \spc
\label{hfold}
\eq
where $\chi = s\,x + (m^2_2 - m^2_1 - s)\,x + m^2_1$. We introduce
\bq
s_{\mrp} = (m_1 - m_2)^2 \spc \quad s_{\mrt} = (m_1 + m_2)^2 \spc \quad
\lambda = (s - s_{\mrp})\,(s - s_{\mrt}) \spc
\eq
and consider the behavior of $f(s)$ around the normal threshold $s_{\mrt}$. We obtain
\bq
f(s) = - 8\,\frac{s^{1/2}}{\lambda} + \;\; \mbox{reg. terms} \qquad s \to s_{\mrt} \spp
\eq
The function has a simple pole at the normal threshold. Inserting this result into \eqn{hfold} we obtain
\bqa
\mrF(s) &=& 8\,\mrH_0\,\beta\,\frac{s^{1/2}}{s_{\mrt} - s_{\mrp}}\,(s - s_{\mrt})^{-1}\,J(s) + \mrF_{\mrp}(s) \spc
\nl
J(s) &=& \int_0^1 dx\,x^{\beta - 1}\,(1 - x)^{1/2}\,(1 + \frac{s}{s_{\mrt} - s}\,x)^{-1} \spp
\eqa
The integral in $J$ gives
\bq
J(s) = \mrB(\frac{3}{2}\,,\,\beta)\,\mrK(s) \spc \quad
\mrK(s) = {}_2\mrF_{1}(1\,,\,\beta\,;\,\beta + \frac{3}{2}\,;\,\frac{s}{s - s_{\mrt}}) \spc
\eq
where ${}_2\mrF_{1}$ is the hypergeometric function and $\mrB$ is the Euler beta function.
For $s \to s_{\mrt}$ we use well-known properties of the hypergeometric function and obtain
\bq
\mrK(s) = \mrB_1\,\rho^{-1}\,{}_2\mrF_{1}(1\,,\,\frac{1}{2} - \beta\,;\,2 - \beta\,;\,\rho^{-1}) +
          \mrB_2\,\rho^{-\beta}\,{}_2\mrF_{1}( - \frac{1}{2}\,,\,\beta\,;\,\beta\,;\,\rho^{-1}) \spc
\eq
where we have introduced
\bq
\rho = \frac{s}{s_{\mrt} - s} \spc \quad
\mrB_1 = \frac{\beta + 1/2}{\beta - 1} \spc \quad
\mrB_2 = 2\,\pi^{-1/2}\,\Gamma(\beta + 3/2)\,\Gamma(1 - \beta) \spp
\eq
Therefore, the leading behavior of $\mrK$ is given by
\bq
\mrK(s) \sim \mrB_2\,s^{-\beta}\,\lpar \frac{s_{\mrt}}{s} \rpar^{1/2}\,(s_{\mrt} - s)^{\beta} \spc
\qquad s \to s_{\mrt} \spp
\eq
In terms of the folding this means that $\mrF(s) \sim (s_{\mrt} - s)^{\beta - 1}$ which is integrable and can be used
in the convolution with the beam energy spread~\footnote{The influence of radiation and energy spread was suggested,
long ago, in a private discussion by Thomas Binoth.}.
When masses become complex, $m^2_i \to m^2_i - i\,m_i\,\Gamma_i$, the zeros of $\lambda$ also become complex
and the solutions of $\Re\,\lambda = 0$ move above $s_{\mrt}$ and below $s_{\mrp}$ by a quantity proportional to
the widths.
\section{AT beyond the SM \label{BSM}}
As mentioned, a physical-region singularity requires a theory with a hierarchy of heavy masses: therefore, ATs 
hardly appear in the SM. However any BSM theory with an heavy neutral Higgs boson ($\PH$) and a charged
one ($\PHpm$) satisfying $M_{\sPH} > 2\,M_{\sPHpm}$ and $M_{\sPHpm} > \mw + \mz$ will have an AT in
the decay $\PH \to \Pl \Pl \PGn \PGn$. 
One could even imagine a situation with a light Higgs boson $\Ph$ and $3$ heavy Higgs bosons, $\PH_{1,2,3}$, where 
$\mrM_{\PH_1} > 2\,\mrM_{\PH_2}$, $\mrM_{\PH_2} > \mrM_{\PH_3} + \mrM_{\Ph}$ and $\mrM_{\PH_3} > 3\,\mrM_{\Ph}$ giving an AT 
in the pentagon corresponding to $\PH_1 \to 6\,\Ph$.

There are also specific examples from a supersymmetric context, namely the production of a heavy neutral Higgs and 
a pair of massless $\PQb\,$-quarks by gluon fusion, via a loop containing two squarks (sbottoms) and two neutralinos;
for instance we mention AT effects in Higgs decays into charginos and neutralinos in the complex 
MSSM~\cite{Heinemeyer:2015pfa}. 

In general the large multiplicity of (super)fields introduced
in non-minimal Susy-GUT models will result in the development of a Landau singularity. 
Another example is given by $M_{\sPHpm} > m_{\PQt} + m_{\PQb}$ in the decay $\PH \to \PAQt + \PQt$.
In general we can say that whenever the initial and final states have more than two particles,
the scattering matrix gets a contribution from diagrams where there are singularities for physical 
values of the energies and momenta~\cite{RevModPhys.33.448}.

\section{Conclusions \label{conc}}
Anomalous thresholds have been studied by physicists since the 1950’s, \eg electromagnetic scattering off a deuteron:
A fast falloff of the form factor of loosely bound deuteron is due to the presence of the anomalous singularity
close to the physical region of the scattering reaction; if not for the anomalous singularity, the scattering 
amplitude would have only vary at a much larger scale of the two-proton threshold.

Causality ensures that scattering amplitudes are analytic functions of momenta and an analytic function is 
characterized by its singularities.
Furthermore, each scattering function has physical-region singularities only on positive-$\,\alpha$
Landau surfaces~\cite{Iagolnitzer:1969sk} and near these surfaces it is the limit from certain well-defined directions of
a unique analytic function.
 
In this work the matrix element corresponding to the general one-loop Feynman diagram is rigorously investigated in 
detail. We have given a general classification of the one-loop - physical-region - leading Landau singularities 
(the so-called anomalous thresholds) for LHC processes, taking into account that the position of each singularity 
is determined by masses and invariants while the character of the singularity derives from the topology of the 
interaction process.

Our methodology is based on finding zeros of Cayley determinants with constraints. The advantage of the approach
is that the condition for an AT is written directly in the space of invariants instead of being obtained through 
a set of consistency equations in $+\,\alpha$ Landau surfaces.

The main motivation was to find observable effects, \eg. peaks in distributions; indeed, resonances are normally 
observed as peaks in certain invariant mass distributions. However, is a peak necessarily due to the presence of 
a resonance? Are there peaks produced by kinematic singularities? 
It is well known that scattering amplitudes also possess singularities corresponding to more
complicated types of particle exchanges: these are indeed the Landau singularities.
A well-know example is that there has been interest for many years in trying to use the triangle
singularity to explain certain enhancements in strongly interacting three-particle final states.

We have found that peaks, when present, are marginal and the main result of our work is that the radiative corrections 
induced by physical-region ATs are well under control once regularized with the complex mass scheme;
nevertheless they should be taken into account in estimating the missing higher order uncertainty~\cite{David:2013gaa}.
When assessing such results, it should be borne in mind that Yukawa suppressed LO processes can be heavily influenced 
by NLO corrections, \ie NLO is the first relevant term. 

We need to acknowledge the fact that we do not have a fully satisfactory (gauge) theory of unstable particles,
despite past~\cite{Veltman:1963th,Weldon:1975gu} and recent progress~\cite{Actis:2006rb,Denner:2014zga}.
However, the complex mass scheme is essential in this context, given the non-integrable character of the ATs for 
boxes and pentagons (if internal masses are kept real). From this point of view we are tempted to argue for a
definition of a ``natural'' theory as the one where there are no ATs inside the physical region. 
To summarize - with anomalous thresholds non-integrable functions may enter into physical calculations, and the attempt 
to interpret these integrals and find useful solutions can lead us to a broader understanding of the physical situation. 

The SM is almost ``natural'': physical-region ATs are exceptional for SM, on-shell, LHC physics 
but more frequent in the so-called ``off-shell'' LHC physics and in BSM models; the reason for that is immediately 
seen in the context of the Coleman{-}Norton~\cite{Coleman:1965xm} and Kershaw~\cite{Kershaw:1971rc} theorems, \ie not 
enough heavy masses in the SM to be in one of the $6(14)$ branches of the physical-region Landau curve for a triangle (box)
diagram. 

We have discussed how the introduction of complex masses gives rise to new configurations, the
\hyperref[Pzero]{Peierls zeros (defined in Eq.(~\ref*{Pzero}))}.
Furthermore, we have shown in \hyperref[Scas]{section \ref*{Scas}} that boxes and pentagons with arbitrary external masses 
may develop non-integrable singularities even when regularized by the complex mass scheme.
 
We have also discussed the folding of anomalous amplitudes, \eg with QED radiation, in
\hyperref[iexa]{section \ref*{iexa}}.

A final result, discussed in \hyperref[nsing]{section \ref*{nsing}}, is the following: 
for $\Pep \Pem$, $\PAQq \PQq$ and $\Pg \Pg$ initial states there are
amplitudes where the initial state couples to a neutral object which can couple to $\PW \PW$ or 
$\PAQt \PQt$. In these cases we can have $\PAQb \PQb \Pg$ and $\PAf \Pf \PGg$ final states producing
a singularity at large $\Pf(\PAf) {-} \Pg(\PGg)$ invariant masses, on top of the more familiar infrared and
collinear ones.

To conclude we can say that the most elementary form of Landau singularities is connected to holomorphic functions 
defined by integrals of rational functions as they appear in the ``perturbative'' approach to quantum field
theory. From a mathematical point of view these holomorphic structures have a deeper meaning and general 
questions (are Landau singularities effectively singular ``somewhere''?) should receive an answer in this context.

\clearpage
\appendix
\section{Technical details \label{TD}}
Production and decay processes are described in terms of Mandelstam invariants where, for a $n \to m$ process,
we have $3\,(n + m) - 10$ independent invariants. 

Consider a process $\mrN \to 0$ with all incoming momenta; a convenient way of describing the boundaries of the phase 
space, the relations between vectors and invariants and the non-linear constraints that arise when $\mrN \ge 6$
id the following. If at least one momentum is such that $p^2_i \not= 0$ we redefine $n$-dimensional momenta and 
introduce components
\bq
p^{\mu}_1 \equiv \lpar {\vec 0}\,,\,m_1\rpar.
\eq
Then we introduce 
\bq
\lpar h^{\ssN} \rpar_{ij} = \spro{p_i}{p_j}, \qquad i,j= 1,\,\cdots\,,\mrN,
\qquad
h^{\ssN\,;\,kl} = h^{\ssN} \qquad 
\mbox{with row}\;k\;\mbox{and column}\;l\;\mbox{removed}.
\eq
The vector $p_2$ is defined by
\bq
p^{\mu}_2 \equiv ,
 \lpar {\vec 0}\,,\,
\sqrt{\frac{\mathrm{det}\,h^2}{\mathrm{det}\,h^1}}\,,\,
-\,\frac{h^2_{12}}{m1} \rpar.
\eq
Next $n$-dimensional vectors will be
\bqa
p^{\mu}_3 &\equiv& 
\lpar {\vec 0}\,,\,
-\,\frac{
         \mathrm{det}\,h^{3\,;\,32}
        }
        {
         \sqrt{
               \mathrm{det}\,h^1\,\mathrm{det}\,h^2
              }
        }\,,\,
\sqrt{\frac{\mathrm{det}\,h^3}{\mathrm{det}\,h^2}}\,,\,
-\,\frac{h^3_{13}}{m_1}\rpar,
\nl
p^{\mu}_4 &\equiv& 
\lpar {\vec 0}\,,\,
-\,
\frac{
      \mathrm{det}\,h^{4\,;\,43\,;\,32}
     }
     {
      \sqrt{
            \mathrm{det}\,h^1\,\mathrm{det}\,h^2
           }
     }\,,\,
\frac{\mathrm{det}\,h^{4\,;\,43}}{\sqrt{\mathrm{det}\,h^2\,\mathrm{det}\,h^3}}\,,\,
\sqrt{\frac{\mathrm{det}\,h^4}{\mathrm{det}\,h^3}}\,,\,
-\,\frac{h^4_{14}}{m_1}\rpar,
\eqa
\etc If $\mrN = 5$ then momentum conservation gives $p_5= -\sum\,p_i$. If, 
instead $\mrN= 6$, $p_6$ follows from momentum conservation whereas $p_5$ 
requires a fifth component; if $\mrd = 4$ the vanishing of the fifth component 
requires $\mathrm{det}\,h^5 = 0$.

Our convention for selecting independent invariants and their boundaries follows \Bref{Kumar:1970cr}.
Loop integrals follow from \Bref{Ferroglia:2002mz}, \ie we perform numerical integration over Feynman
parameters and Mandelstam invariants in one stroke. It is worth noting that instabilities due to zeros of 
Gram determinants are absent in this approach. The typical integrals to be evaluated are of the following form:
\bq
\mrO_{\alpha} = \int d\{\mrI\}\,\dsimp{\mrN}\,\sum_{j_1=0}^{\mrJ_1}\,\dots\,\sum_{j_{\ssN}=0}^{\mrJ_{\ssN}}\,
\mrF_{j_1\,\dots\,j_{\ssN}}(\{\mrI\})\,x^{j_1}_1\,\dots\,x^{j_{\ssN}}_{\ssN}\,
\Bigl[ \lpar x - X_{\ssN+1} \rpar^t\,\mrH\,\lpar x - X_{\ssN+1} \rpar + \Delta_{\ssN+1} \Bigr]^{-\alpha} \spc
\eq
where $\{I\}$ denotes the set of Mandelstam invariants.

Kershaw theorem proves factorization of the scattering amplitude in the vicinity of the given Landau singularity
but also that, for a given set of invariants which lie on the given physical-region Landau singularity, the loop
momentum is uniquely determined. This means that, in the vicinity of the singularity, all loop integrals are
scalar. In Feynman parameter space this can be seen as follows:
\bq
\mrO_{\alpha}\bmid_{\AT} \sim \int d\{\mrI\}\,
\sum_{j_1=0}^{\mrJ_1}\,\dots\,\sum_{j_{\ssN}=0}^{\mrJ_{\ssN}}\,
\mrF_{j_1\,\dots\,j_{\ssN}}(\{\mrI\})\,X^{j_1}_1\,\dots\,X^{j_{\ssN}}_{\ssN}\,
\dsimp{\mrN}\,
\Bigl[ \lpar x - X_{\ssN+1} \rpar^t\,\mrH\,\lpar x - X_{\ssN+1} \rpar + \Delta_{\ssN+1} \Bigr]^{-\alpha} \spp
\eq
\subsection{The phase-space integral for \texorpdfstring{ $\Pg \Pg  \to  \PAQb \PQb \PX$}{} \label{fps}}
Consider the process 
\bq
\Pg(p_1) + \Pg(p_2) \to \PQb(q_1) + \PX(q_2) + \PAQb(q_3) \spc
\eq
with $s = - (p_1 + p_2)^2$, $p^2_i = 0$, $q^2_{1,3} = - m^2_{\PQb}$ and $q^2_2 = \mrM^2$. We introduce
\bq
\lambda_1 = \lambda(s\,,\,s_1\,,\,m^2_{\PQb}) \spc \quad
\lambda_2 = \lambda(s_1\,,\,\mrM^2\,,\,m^2_{\PQb}) \spc \quad
\lambda_3 = \lambda(s\,,\,u_1\,,\,\mrM^2)
\eq
and auxiliary quantities
\bqa
\xi &=& \lambda^{-1/2}_1\,(s - s_1 + 2\,t_0 - m^2_{\PQb}) \spc
\nl
\eta &=& (\lambda_1\,\lambda_3)^{-1/2}\,\Bigl[ 2\,s\,(s_1 + \mrM^2 - m^2_{\PQb}) -
         (s - u_1 +\mrM^2)\,(s + s_1 - m^2_{\PQb}) \Bigr] \spc
\nl
\omega &=& \lambda^{-1/2}_3\,(s - u_1 + 2\,t_ - \mrM^2) \spc
\nl
\zeta &=& (1 - \eta^2)^{-1/2}\,(1 - \xi^2)^{-1/2}\,(\omega - \xi \eta) \spp
\eqa
The corresponding phase-space integral is
\bq
\int d\Phi = \frac{\pi}{2}\,
\int_{s_{1-}}^{s_{1+}} ds_1\,
\int_{u_{1-}}^{u_{1+}} du_1\,
\int_{t_{0-}}^{t_{0+}} dt_0\,
\int_{t_{1-}}^{t_{1+}} dt_1\,
(\lambda_1\,\lambda_3)^{-1/2}\,\Bigl[ (1 - \eta^2)\,(1 - \xi^2)\,(1 - \zeta^2) \Bigr]^{-1/2} \spp
\eq
The boundaries are 
\bq
s_{1-} = (\mrM + m_{\PQb})^2 \spc \quad
s_{1+} = (\sqrt{s} - m_{\PQb})^2 \spc \quad
u_{1\pm} = u_{10} \pm \Delta u_1 \spc \quad \mbox{\etc}
\eq
\bqa
u_{10} &=& s + \mrM^2 - \frac{1}{2\,s_1}\,(s_1 + \mrM^2 - m^2_{\PQb})\,(s + s_1 - m^2_{\PQb}) \spc
\nl
t_{00} &=& m^2_{\PQb} - \frac{1}{2}\,(s - s_1 + m^2_{\PQb}) \spc
\nl
t_{10} &=& \mrM^2 - \frac{1}{2}\,(s - u_1 + \mrM^2) + \frac{1}{2}\,\lambda^{1/2}\,\xi\,\eta \spc
\eqa
\bq
\Delta u_1 = \frac{1}{2}\,\frac{(\lambda_1\,\lambda_2)^{1/2}}{s_1} \spc \quad
\Delta t_0 = \frac{1}{2}\,\lambda^{1/2}_1 \spc \quad
\Delta t_1 = \frac{1}{2}\,\lambda^{1/2}_3\,\Bigl[ (1 - \eta^2)\,(1 - \xi^2) \Bigr]^{1/2} \spp
\eq
Introducing dimensionless variables,
\bq
s_1 = (s_{1+} - s_{1-})\,y_1 + s_{1-} \spc \qquad
u_1 = 2\,\Delta u_1\,y_2 + u_{1-} \spc \qquad \mbox{etc} \spc
\eq
we obtain
\bq
\int d\Phi = \frac{\pi}{2}\,(s_{1+} - s_{1-})\,\int\,\prod_{i=1}^{4}\,dy_i\,
\frac{(\lambda_1\,\lambda_2)^{1/2}}{s_1\,(1 - \zeta^2)^{1/2}} \spp
\eq
Alternatively, we observe that $\zeta(t_{1\pm}) = \pm 1$ and
\bq
\int_{t_{1-}}^{t_{1+}}\,dt_1 = \frac{1}{2}\,\int_{-1}^{+1}\,d\zeta\,
\Bigl[ \lambda_3\,(1 - \xi^2)\,(1 - \eta^2) \Bigr]^{1/2} \spc
\eq
introduce
\bq
\zeta = \sin \mrz \spc \qquad
\mrz = \pi\,y_4 - \frac{1}{2}\,\pi \spc
\eq
and derive
\bq
\int d\Phi = \frac{\pi^2}{4}\,(s_{1+} - s_{1-})\,\int \prod_{i=1}^{4}\,dy_i\,
\frac{(\lambda_1\,\lambda_2)^{1/2}}{s_1} \spp
\eq
In this way the phase-space integral is mapped into the unit, four-dimensional, cube.
\subsection{Momenta and invariants: an example}
Consider the process 
\bq
\Pg(p_1) + \Pg(p_2) \to \PQb(q_1) + \PH(q_2) + \PH(q_3) + \PAQb(q_4) \spc
\label{exapro}
\eq
momentum conservation gives $q_4 = P - q_1 - q_2 - q_3$ where $P = p_1 + p_2$. There are 8 invariants~\footnote{For $n \to m$
there are $3\,(n + m) - 10$ invariants.} defined by~\cite{Kumar:1970cr}:
\bqa
s &=& - (p_1 + p_2)^2 \spc
\nl
s_i &=& - \lpar P - \sum_{j=1}^{i}\,q_i \rpar^2 \spc \qquad i=1,2 \spc
\nl
u_i &=& - \lpar P - q_{i+1} \rpar^2 \spc \qquad i=1,2 \spc
\nl
t_i &=& - \lpar p_1 - q_{i+1} \rpar^2 \spc \qquad i=0,1,2 \spp
\eqa
The linear relations among scalar products, $\mrT_{ij} = \spro{p_i}{q_j}$, $\mrS_{ij} = \spro{q_i}{q_j}$  
and invariants are as follows:
{\footnotesize{
\[
\begin{array}{llll}
\toprule
\hline
&&&\\
\mrT_{11} = \frac{1}{2}\,(t_0 - m^2_{\PQb})\quad & \quad
\mrT_{12} = \frac{1}{2}\,(t_1 - \mhs)                               \quad & \quad
\mrT_{13} = \frac{1}{2}\,(t_2 - \mhs)                               \quad & \quad
\mrT_{14} = \frac{1}{2}\,(m^2_{\PQb} - t_0 - t_1 - t_2 - s) + \mhs                               \\
&&&\\
\hline
&&&\\
\mrT_{21} = \frac{1}{2}\,(s_1 - t_0 - s)      \quad & \quad
\mrT_{22} = \frac{1}{2}\,(u_1 - t_1 - s)                                 \quad & \quad
\mrT_{23} = \frac{1}{2}\,(u_2 - t_2 - s) \quad & \quad 
\mrT_{24} = \frac{1}{2}\,(t_0 + t_1 + t_2 - u_1 - u_2 - s_1) +s                                 \\
&&&\\
\hline
&&&\\
\mrS_{12} = \frac{1}{2}\,(s_1 - s_2 + u_1 - s)                                            \quad & \quad
\mrS_{13} =\frac{1}{2}\,(s_2 + u_2 - s - m^2_{\PQb}) - \mrS_{23} \quad & \quad 
\mrS_{14} = \frac{1}{2}\,(s - u_1 - u_2) + m^2_{\PQb} + \mrS_{23}                      \quad & \quad
                                                                                                \\
&&&\\
\hline
&&&\\
\mrS_{24} = \frac{1}{2}\,(s_2 - s_1 + \mhs) - \mrS_{23}           \quad & \quad
\mrS_{34} = \frac{1}{2}\,(\mhs + m^2_{\PQb} - s_2)                      \quad & \quad
                                                                              \quad & \quad
                                                                              \\
&&&\\
\hline
\bottomrule
\end{array}
\]
}}
\begin{center}
Table A2: The linear relations among scalar products for process \eqn{exapro}.
\end{center}

One scalar product remains free. To fix it we proceed as follows:
\begin{enumerate}

\item Go to the c.m.s with $p_1$ and $p_2$ along the $\mrz\,$-axis and $\spro{p_1}{p_2} = - s/2$.

\item Put $q_1$ in the $\mrx-\mrz$ plane and fix its components (with $q^2_1 = - m^2_{\PQb}$) in terms
of $\spro{p_1}{q_1}$ and $\spro{p_2}{q_2}$.

\item Add the $\mry\,$-component for $q_2$ and fix $q_2$ (with $q^2_2= - \mhs$) in terms of the scalar
products between $q_2$ and $p_1, p_2, q_1$.

\item Introduce a fifth component for $q_3$, derive $q_3$ (with $q^2_3 = - \mhs$) in terms of the scalar
products between $q_3$ and $p_1, p_2, q_1, q_2$.

\item Replace scalar products in terms of invariants. The equation requiring this fifth component to be zero
is a non-linear constraint, \ie a quadratic equation in $\mrS_{23}$ with coefficients depending on
the eight linearly independent invariants $s,\,s_1,\,\dots\,,t_2$.

\end{enumerate}
If we consider the subprocess
\bq
\Pg + \Pg \to \PQb + \PAQb + \PH (\to \PH \PH) \spc
\eq
there is no need to introduce the full $2 \to 4$ kinematics and the phase space is a simple convolution of
$2 \to 3$ and $1 \to 2$.
\section{One-loop integrals around their AT  \label {SB}}

In this appendix we present explicit results for the leading behavior of one-loop integrals around their physical-region AT; 
the original derivation was given in \Bref{Ferroglia:2002mz}.

To extract the leading behavior of a triangle around its leading Landau singularity we introduce
\bq
\mrC_{\Box}\lpar p_1,p_2\,;\,m_1,m_2,m_3\rpar =
\mrC_0\lpar p_1,p_2\,;\,m_1,m_2,m_3\rpar +
\mrC_0\lpar p_2,p_1\,;\,m_3,m_2,m_1\rpar \spp
\label{Cbox}
\eq
If the first $\mrC_0$ in the r.h.s. of \eqn{Cbox} is singular then the second is regular and
\bq
\mrC_0\lpar p_1,p_2\,;\,m_1,m_2,m_3\rpar \sim 
\mrC_{\Box}\lpar p_1,p_2\,;\,m_1,m_2,m_3\rpar \spp
\eq
Given the quadratic forms
\bq
\mrV_3\lpar x_1,x_2\rpar = x^t\,\mrH\,x + 2\,\mrK^t\,x + \mrL \spc
\eq
\bq
\mrV^{(1)}(x) = \mrV_3(0,x) \spc \quad 
\mrV^{(2)}(x) = V_3(x,0) \spc \quad 
{\overline \mrV}^{(1)}(x)= \mrV_3(1,x) \spc \quad
{\overline \mrV}^{(2)}(x)= \mrV_3(x,1) \spc
\eq
consider two-dimensional bubbles
\bq
\mrB^{(i)}_2 = \int_0^1 dx\,\frac{1}{\mrV^{(i)}(x)} \spc \qquad
{\overline \mrB}^{(i)}_2 = \int_0^1 dx\,\frac{1}{{\overline \mrV}^{(i)}(x)}.
\eq
We find
\bq
\mrC_{\Box} \sim -\,\frac{1}{2}\,\ln \Delta_3\,\sum_{i=1,2}\,\Bigl[
X_i\,\mrB^{(i)}_2 + \lpar 1 - X_i\rpar\,{\overline \mrB}^{(i)}_2 \Bigl] \spc
\label{C0beh}
\eq
for $\Delta_3 \to 0$. 
For generalized triangles we obtain
\bq
\mrC_{\Box}(i,j) \sim X^i_1\,X^j_2\,\mrC_{\Box} \spc
\eq
as expected by the fact that the AT is a pinch singularity (and by the Kershaw theorem).

To extract the leading behavior of a box around its leading Landau singularity we introduce
\bq
\mrD_{\Box} = \sum_{\{123\}}\,\dsimp{3}\,\mrV^{-2}_4\lpar x_1,x_2,x_3\rpar \spc
\eq
where the sum is over the permutations of $x_1,x_2,x_3$. 
The first term in the sum is the original $\mrD_0$ function while the rest gives 
the five complementary functions which, by construction, are regular at the
AT. As a consequence, we now have to evaluate $\mrD^{\Box}_0$ when $\Delta_4 
\approx 0$ and the point of coordinates $X_i$ is inside the unit cube.
We introduce
\bq
\mrV^{(1)}\lpar x_1,x_2\rpar = \mrV_4\lpar 0,x_1,x_2\rpar \spc
\quad
\mrV^{(2)}\lpar x_1,x_2\rpar = \mrV_4\lpar x_1,0,x_2\rpar \spc
\quad
\mrV^{(3)}\lpar x_1,x_2\rpar = \mrV_4\lpar x_1,x_2,0\rpar \spc
\eq
\bq
{\overline \mrV}^{(1)}\lpar x_1,x_2\rpar = \mrV_4\lpar 1,x_1,x_2\rpar \spc
\quad
{\overline \mrV}^{(2)}\lpar x_1,x_2\rpar = \mrV_4\lpar x_1,1,x_2\rpar \spc
\quad
{\overline \mrV}^{(3)}\lpar x_1,x_2\rpar = \mrV_4\lpar x_1,x_2,1\rpar \spc
\eq
where we have put $\Delta_4 = 0$, and consider the $3\,$-dimensional $\Box\,$-triangles
\bq
\mrC^{(i)}_3 = \int_0^1 dx_1 \int_0^{x_1} dx_2\,
\Bigl[ \mrV^{(i)}\lpar x_1,x_2\rpar\Bigr]^{-3/2} \spc \qquad
{\overline \mrC}^{(i)}_3 = \int_0^1 dx_1 \int_0^{x_1} dx_2\,
\Bigl[ {\overline \mrV}^{(i)}\lpar x_1,x_2\rpar\Bigr]^{-3/2} \spp.
\eq
The result is
\bq
\mrD_{\Box} \sim \egam{\frac{1}{2}}\,\Delta^{-1/2}_4\,\sum_{i=1,3}\,\Bigl[
X_i\,\mrC^{(i)}_3 + \lpar 1 - X_i\rpar\,{\overline \mrC}^{(i)}_3 \Bigr] \spp
\label{D0beh}
\eq
The results of \eqns{C0beh}{D0beh} have been derived under the assumption that the corresponding Gram determinant is 
not vanishing. If this is not the case we will write the box(triangle) as a linear combination of four(three)
triangles(bubbles) divided by the corresponding Cayley determinant~\cite{Ferroglia:2002mz}. An example will help in
understanding; consider the following integral:
\bq
\mrI = \int_0^1 dx\,\int_0^{x} dy \Bigl[ x^2 - \lambda\,y^2 + 2\,(a\,x - \lambda\,b\,y) + \mrL \Bigr]^{-1} \spp
\eq
In this case we derive
\[
\mrH = \left(
\begin{array}{cc}
1   & 0 \\
0 & \lambda \\
\end{array}
\right)
\]
with $X_1 = - a, X_2 = b$, $\mrG = \lambda$ and $\mrC = \mrL - a^2 - \lambda\,b^2$. We assume that 
$0 \le b \le - a \le 1$ and derive the usual result
\bq
\mrI \sim \ln \frac{\mrC}{\mrG} \spc \qquad \Delta = \frac{\mrC}{\mrG} \to 0 \spp
\eq
However, if we take the limit $\lambda \to 0$ first then $\mrG = 0$. In this case we obtain
\bq
\mrI\mid_{\lambda = 0} \sim ( \mrL - a^2 )^{-1/2} \spc \qquad \mrL \to a^2 \spc
\eq
where $\mrL - a^2$ is the Cayley determinant evaluated at $\mrG = 0$. Therefore, the behavior is $\mrC^{-1/2}$ and
not $\ln \Delta$. 
 
 \clearpage
\bibliographystyle{elsarticle-num}
\bibliography{PaC}

\end{document}